\documentclass[buildings]{Definitions/mdpi} 
\firstpage{1} 
\pubvolume{1}
\issuenum{1}
\articlenumber{0}
\pubyear{2026}
\copyrightyear{2026}

\datereceived{ } 
\daterevised{ } 
\dateaccepted{ } 
\datepublished{ } 

\Title{AffectCity: An Empirical Investigation of Complexity, Transparency, and Materiality in Shaping Affective Perception of Building Facades}

\Author{Chenxi Wang $^{1}$, Haining Ding $^{2}$ and Michal Gath-Morad $^{2,}$*}
\AuthorNames{Chenxi Wang, Haining Ding and Michal Gath-Morad}

\address{%
$^{1}$ \quad Cambridge Cognitive Architecture, Department of Architecture, University of Cambridge, UK\\
$^{2}$ \quad NeuroCivitas Lab for NeuroArchitecture, Centre for Research in the Arts, Social Sciences and Humanities (CRASSH), University of Cambridge, UK}

\corres{Correspondence: mg2068@cam.ac.uk}

\abstract{Buildings shape how people feel, yet the mechanisms 
through which specific fa\c{c}ade properties drive 
affective states remain empirically underspecified. 
Here we introduce the Cambridge Fa\c{c}ade Affect 
Dataset (CFAD), 86 orthogonally rectified fa\c{c}ade 
images annotated with continuous arousal and valence 
ratings from 85 participants, and establish a 
validated pipeline linking machine-vision-derived 
surface metrics to human affective responses. 
Focusing on three quantifiable attributes, 
\textit{complexity}, \textit{transparency} 
(window-to-wall ratio), and \textit{materiality} 
(proportion of natural versus artificial surface 
composition), we show that perceived complexity is 
the dominant affective predictor, with significant 
positive associations for both arousal 
(\texorpdfstring{$\beta = 0.507$}{beta = 0.507}, 
\texorpdfstring{$p < 0.001$}{p < 0.001}) and valence (\texorpdfstring{$\beta = 0.376$}{beta = 0.376}, 
\texorpdfstring{$p < 0.001$}{p < 0.001}) and a curvilinear 
amplification at higher complexity levels. 
Transparency exhibits an inverted-U relationship 
with valence, while increasing surface artificiality 
suppresses arousal and reduces pleasantness 
consistent with biophilic response theory. 
Critically, machine-derived metrics show limited 
direct predictive power over affective outcomes; 
mediation analyses reveal that human perceptual 
evaluation functions as a necessary intermediate 
layer, with perceived materiality significantly 
mediating the machine--valence relationship 
(indirect effect $= -0.205$, $p = 0.003$). 
Cross-context validation demonstrates moderate 
stability of complexity and materiality ratings 
across image-based and in-situ conditions, while 
affective responses, particularly valence, exhibit 
significant context-dependence (ICC $= 0.332$). 
These findings advance fa\c{c}ade research from 
descriptive morphological analysis toward 
predictive, perception-grounded modelling, and 
provide an empirically validated basis for 
affect-informed design of the urban environment.}

\keyword{architectural fa\c{c}ades; affective 
perception; arousal; valence; computer vision; 
neuroarchitecture; urban affect; built environment}

\begin{document}
\hypersetup{pdfencoding=auto}
\setlength{\headheight}{20.0pt}
\addtolength{\topmargin}{-8.0pt}

\section{Introduction}
\subsection{Problem Statement}

Urban environments represent one of the most pervasive yet 
incompletely understood determinants of human psychological 
experience. While substantial evidence links macro-scale features 
of the built environment, including green infrastructure, street 
network configuration, and spatial density, to neural stress 
responses and affective wellbeing \citep{Lederbogen2011, Tost2015, 
Nieuwenhuijsen2021}, the building fa\c{c}ade, the most immediate 
and continuously encountered surface of the urban fabric, 
remains comparatively underexplored as a primary driver of 
affective experience. This gap is scientifically consequential. 
Human visual processing is selectively tuned to structural 
regularities encoded within architectural surfaces, including 
edge density, bilateral symmetry, material texture, and spatial 
rhythm, features that engage the sensory-motor, emotion-valuation, 
and meaning-knowledge neural systems implicated in aesthetic and 
affective appraisal \citep{Chatterjee2014, Vartanian2013, 
Coburn2017, Wang2025} Survey data reinforce this perceptual relevance, 
with 76\% of UK residents associating building appearance with 
shifts in emotional state \citep{Heatherwick2023}, yet the 
mechanisms through which specific fa\c{c}ade properties translate 
into quantifiable affective responses have received limited 
empirical attention.

Affect, understood through the orthogonal dimensions of 
\textit{arousal} and \textit{valence}, provides a 
neurobiologically grounded and dimensionally tractable framework 
for characterising environmental quality \citep{Russell1980, 
RussellBarrett1999, Barrett2017}. The emerging field of 
environmental neuroscience has begun to map the neural pathways 
through which physical settings modulate emotional and mental 
health outcomes, positioning specific properties of the built 
environment, rather than broad urbanicity per se, as candidate 
active ingredients of psychological experience \citep{Kuhn2024, 
Pessoa2008}. This perspective is increasingly reflected in 
global health policy, with the WHO Healthy Cities programme 
recognising affectively supportive environments as a public 
health priority \citep{WHO2016, Nieuwenhuijsen2021}. Within 
the design disciplines, it has catalysed the growth of 
neuroarchitecture as an empirical research programme 
\citep{Coburn2017, Coburn2020}, embodied in dedicated 
institutional infrastructure including the Cambridge Cognitive 
Architecture Lab, the MIT Senseable City Lab, and the UCL 
Centre for Neuroarchitecture and Neurodesign, alongside 
practice-led initiatives such as Heatherwick Studio's 
``Humanise'' campaign \citep{Heatherwick2023}. Despite this 
convergence of scientific and design interest, empirical 
progress has been constrained by two structural methodological 
limitations.

First, existing computational analyses of urban perception 
operate predominantly at the streetscape scale, treating 
fa\c{c}ades as one element among many within aggregated visual 
indices. Studies quantifying walkability \citep{EwingHandy2009}, 
visual appeal \citep{Dubey2016}, and perceived scenicness 
\citep{Seresinhe2017} demonstrate that built environment 
properties influence human judgement, yet their composite 
units of analysis preclude identification of the independent 
affective contribution of the building envelope. This matters 
empirically: architectural interiors elicit distinct, 
replicable patterns of neural activity across coherence, 
fascination, and hominess dimensions \citep{Coburn2020}, 
while contour properties alone modulate approach-avoidance 
decisions in architectural contexts \citep{Vartanian2013}. 
At the fa\c{c}ade level specifically, the affective 
consequences of surface articulation, visual complexity, and 
material composition remain poorly characterised in controlled 
experimental settings, and their mapping onto dimensional 
affective outcomes, arousal and valence, has yet to be 
established.

Second, advances in computational image analysis have produced 
increasingly precise quantification of fa\c{c}ade properties, 
including edge density \citep{Marr1980}, texture descriptors 
\citep{Doersch2012}, and fractal dimension \citep{Stamps2002}, 
yet these metrics have received limited empirical validation 
against human affective responses. Studies of fa\c{c}ade 
appearance, including those examining blank versus articulated 
surfaces \citep{WangMunakata2024} and dynamic adaptive skins 
\citep{Beatini2024}, tend to rely on subjective preference 
ratings under short-term or ecologically limited exposure 
conditions, without integration into established dimensional 
affective frameworks \citep{Nasar1994, Akalin2009}. 
Consequently, a structural disconnect persists between 
formal geometric description and perceptual experience: 
computational descriptors may capture properties of physical 
structure that are only partially, or indirectly, related to 
the affective states they are assumed to elicit 
\citep{Gregorians2022}.

Here, we address this dual limitation by introducing the 
Cambridge Fa\c{c}ade Affect Dataset (CFAD), comprising 86 
orthogonally rectified fa\c{c}ade images from Cambridge, UK, 
annotated with continuous arousal and valence ratings from 
85 participants. We focus on three quantifiable fa\c{c}ade 
attributes, \textit{complexity}, \textit{transparency} 
(window-to-wall ratio), and \textit{materiality} (proportion 
of natural versus artificial surface), and establish an 
integrated analytical pipeline linking machine-vision-derived 
metrics to human perceptual evaluations. Through two 
complementary validation stages, one assessing perceptual 
alignment between computational and human-rated attributes, 
and one testing ecological validity across online and 
field-based exposure conditions, we demonstrate that 
specific fa\c{c}ade properties systematically predict 
affective responses. These findings advance façade research 
from descriptive morphological analysis toward predictive, 
perception-grounded modelling, and offer an empirically 
validated framework for affect-informed architectural design.

\subsection{Research Gaps and Questions}

\begin{figure}[H]
\includegraphics[width=\textwidth]{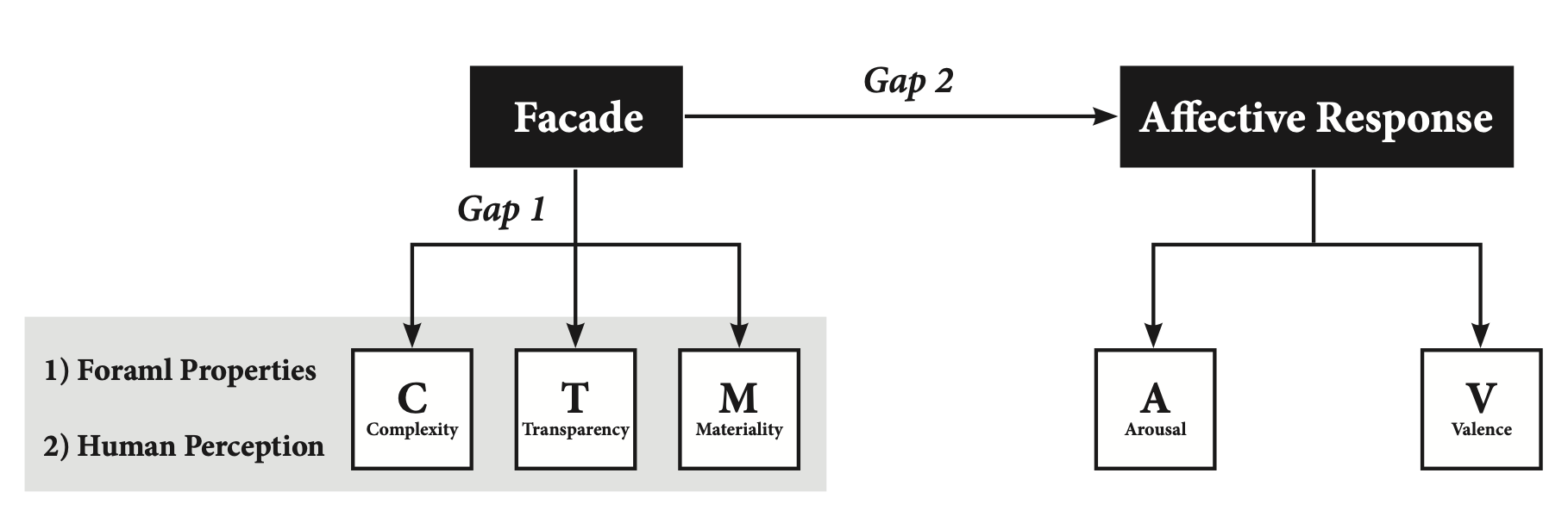}
\caption{Pipeline linking fa\c{c}ade features and 
affective responses. This diagram illustrates the 
dual pathway from architectural fa\c{c}ades to 
affective experience via formal properties 
(complexity, transparency, materiality: \textit{C}, 
\textit{T}, \textit{M}) and human perception towards 
emotional outcomes (arousal and valence: \textit{A}, 
\textit{V}). Two critical gaps (Gap 1, Gap 2) 
highlight underexplored transitions between physical 
form, subjective appraisal, and affective response. 
Source: Author.\label{fig1}}
\end{figure}
\unskip
\vspace{6pt}

Two structural gaps in the current literature motivate the 
present study. The first concerns the unit of analysis. 
Existing research on urban affective experience has operated 
predominantly at the scale of composite streetscapes, 
treating the building envelope as one element among many 
within aggregated perceptual indices. Influential frameworks 
quantifying walkability \citep{EwingHandy2009}, safety and 
visual appeal \citep{Dubey2016}, and perceived scenicness 
\citep{Seresinhe2017} demonstrate that built environment 
properties shape human judgement, yet their composite 
units of analysis preclude isolation of the independent 
perceptual and affective contribution of the fa{\c{c}}ade 
itself. As a consequence, fa{\c{c}}ade-level morphological 
variables remain unisolated in the empirical literature, 
and the causal or predictive relationships between specific 
surface properties and dimensional affective outcomes remain 
difficult to establish \citep{Gregorians2022, Lindal2013}.

The second gap concerns empirical validation. Advances in 
computational image analysis have yielded increasingly 
precise representations of fa{\c{c}}ade structure, including 
edge density \citep{Marr1980}, texture descriptors 
\citep{Doersch2012}, and fractal dimension 
\citep{Stamps2002}. Yet these metrics have received 
limited systematic validation against human affective 
responses. Computational descriptors are frequently treated 
as proxies for perceptual experience without empirical 
verification, and tend to be applied under uncontrolled or 
ecologically indirect conditions rather than standardised, 
stimulus-controlled evaluations of individual fa{\c{c}}ades 
\citep{Nasar1994, Akalin2009}. The result is a persistent 
misalignment between formal geometric characterisation and 
the affective states such characterisation is assumed to 
index. Addressing both gaps requires a framework that 
simultaneously isolates fa{\c{c}}ade-level features, 
quantifies them through validated computational pipelines, 
and anchors their measurement to empirically grounded 
affective outcomes.

\subsection{Research Design}
\begin{figure}[H]
\begin{adjustwidth}{-\extralength}{0cm}
\centering
\includegraphics[width=\linewidth]{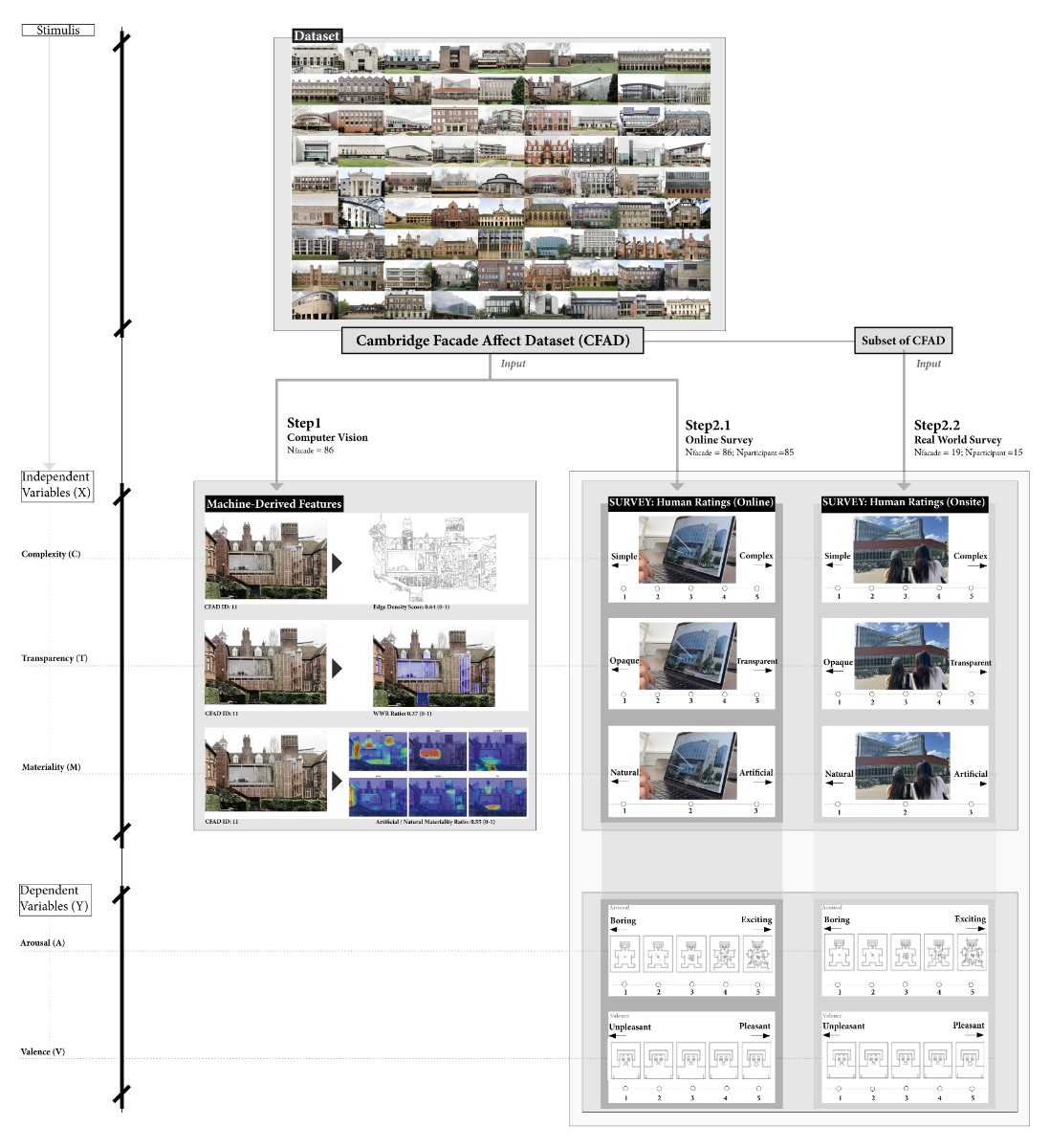}
\caption{Research design of the Cambridge 
Fa\c{c}ade Affect Dataset (CFAD). This schematic 
outlines the multi-stage construction of the CFAD, 
integrating machine-derived fa\c{c}ade features 
(\textit{C}/\textit{T}/\textit{M}: complexity, 
transparency, materiality) with human-rated 
perceptual and affective responses (\textit{A}/\textit{V}: 
arousal, valence). The dataset comprises an 
86-fa\c{c}ade corpus processed via computer vision 
(Step 1), rated by 85 participants in an online 
survey (Step 2.1), and validated through a 
field-based survey with a subsample of 19 
participants across 15 fa\c{c}ades (Step 2.2). 
Source: Author.\label{fig2}}
\end{adjustwidth}
\end{figure}
\unskip
\vspace{6pt}
Here we introduce the Cambridge Fa{\c{c}}ade Affect Dataset 
(CFAD; $N_{\mathrm{f}} = 86$ fa{\c{c}}ades, 
$N_{\mathrm{p}} = 85$ participants), a controlled image-based 
dataset of orthogonally rectified building fa{\c{c}}ades 
from Cambridge, UK, annotated with continuous arousal and 
valence ratings. We focus on three quantifiable fa{\c{c}}ade 
attributes: \textit{complexity}, \textit{transparency} 
(window-to-wall ratio), and \textit{materiality} (proportion 
of natural versus artificial surface composition). For each 
attribute, machine-vision-derived metrics are computed and 
paired with human perceptual ratings, enabling systematic 
examination of the degree to which computational descriptors 
capture perceptually salient properties. Two validation 
stages are implemented: the first quantifies perceptual 
alignment between machine-extracted and human-rated 
fa{\c{c}}ade attributes across the online sample; the 
second evaluates ecological validity by comparing online 
ratings with those obtained from a field-based survey 
($N_{\mathrm{p}} = 19$), testing the stability of affective 
judgements under real-world exposure conditions. Following 
validation, three inferential pathways model the 
relationships between fa{\c{c}}ade attributes and affective 
outcomes: linking human-rated attributes to arousal and 
valence, linking machine-derived metrics to affective 
ratings, and comparing affective responses across online 
and in-situ conditions.

\subsection{Research Hypotheses}
Drawing on information-theoretic accounts of visual 
complexity \citep{Berlyne1971, Imamoglu2000}, the 
affordance properties of transparency \citep{Gibson1979, 
Lindal2013}, and the restorative and stress-buffering 
effects of natural materials \citep{Kaplan1995, Ulrich1991}, 
we hypothesise that: (H1) fa{\c{c}}ade complexity is 
positively associated with arousal and exhibits an inverted 
U-shaped relationship with valence, with pleasantness 
peaking at intermediate levels of visual information density; 
(H2) transparency has a curvilinear relationship with 
arousal and a positive association with valence, reflecting 
the psychophysiological benefits of visual access and 
spatial openness; and (H3) the proportion of natural 
materials is positively associated with both arousal, up 
to moderate-to-high levels, and valence, consistent with biophilic and restorative theories of environmental 
preference. Together, these hypotheses position the 
fa{\c{c}}ade as a tractable and theoretically grounded 
unit of affective inquiry, and the present study as a 
first step toward predictive, perception-grounded modelling 
in affect-informed architectural design.

\subsection{Theoretical Framework: Affect, Architectural 
Perception, and Fa\c{c}ade Morphology}

\subsubsection{Arousal and Valence as Operational Dimensions 
of Environmental Affect}

Human affective responses to the built environment are most 
parsimoniously characterised through two orthogonal dimensions: 
\textit{valence}, reflecting the hedonic tone of an experience 
along a pleasant-to-unpleasant continuum, and \textit{arousal}, 
reflecting physiological and psychological activation 
\citep{Russell1980, RussellBarrett1999}. Originally formalised 
within cognitive psychology, the circumplex model of affect 
has been widely adopted as an operational framework in 
environmental psychology and neuroarchitecture, providing a 
dimensional space into which affective responses to diverse 
stimuli, including architectural and urban settings, can be 
systematically mapped \citep{Barrett2017, Gregorians2022}.

\begin{figure}[H]
\includegraphics[width=\textwidth]{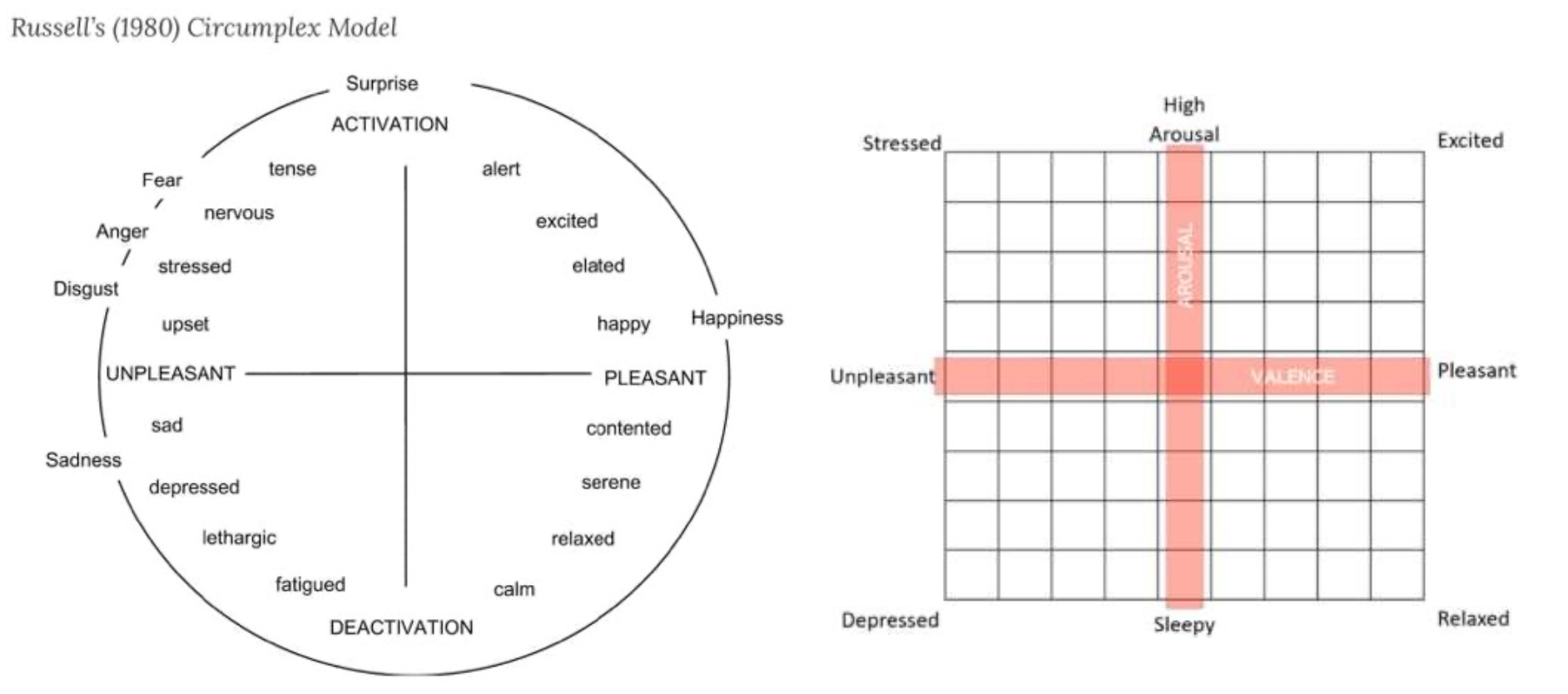}
\caption{Russell's Circumplex Model of Affect and 
modified affect grid. Russell's \citep{Russell1980} 
Circumplex Model illustrates core affect along two 
orthogonal dimensions: valence (pleasure--displeasure) 
and arousal (activation--deactivation). A modified 
affect grid based on this model was adapted for 
study purposes to measure emotional responses to 
architectural fa\c{c}ades. Source: 
\url{https://psu.pb.unizin.org/psych425/chapter/circumplex-models/}\label{fig6}}
\end{figure}
\unskip
\vspace{6pt}

The process through which fa\c{c}ade attributes generate 
affective responses can be understood as a layered 
perceptual-evaluative sequence. Initial sensory registration 
of physical surface properties, including colour, texture, and 
geometric organisation, engages low-level visual processing 
pathways sensitive to edge structure, spatial frequency, and 
material reflectance \citep{Gibson1979, Marr1980}. These 
signals are subsequently integrated with memory, expectation, 
and contextual knowledge through higher-order cognitive systems 
\citep{Neisser1967}. Crucially, affective reactions to visual 
stimuli frequently arise prior to and independently of 
deliberate appraisal: rapid pre-attentive responses to 
symmetry, rhythmic pattern, and material texture can generate 
measurable arousal and valence shifts before conscious 
evaluation is engaged \citep{Zajonc1984, Pessoa2008}. A 
secondary appraisal layer then integrates these initial 
responses with associative and contextual factors, producing 
more elaborated environmental attributions such as perceived 
safety, invitation, or oppressiveness \citep{Lazarus1991, 
Scherer2005}. This cascade from sensory registration through 
affective appraisal constitutes the mechanism by which 
fa\c{c}ade properties are translated into dimensional 
emotional states.

Recent empirical work has extended this framework to the 
built environment. Physiological studies employing EEG and 
skin conductance demonstrate that architectural and urban 
stimuli generate reliable, stimulus-dependent patterns within 
the valence-arousal space \citep{Aspinall2013, Zhang2021}. 
At the scale of building surfaces specifically, 
\citet{Gregorians2022} show that first-person-view exposure 
to architectural scenes produces coherent affective profiles 
along valence and arousal dimensions, with distinct surface 
configurations yielding statistically separable emotional 
responses. These findings position the circumplex model not 
merely as a descriptive psychological framework, but as a 
tractable operational interface between measurable 
architectural attributes and quantifiable affective states.

\begin{table}[H]
\caption{Empirical predictors of affective response 
to fa\c{c}ade attributes. Summary of key 
fa\c{c}ade-related predictors influencing affective 
responses (valence and arousal), categorised by 
physical property: complexity, transparency, 
materiality, and environmental context. Each 
sub-variable is supported by peer-reviewed empirical 
evidence, with significance levels indicated 
($p < .05$, $p < .01$, $p < .001$). This synthesis 
guides the operationalisation of fa\c{c}ade 
attributes in the current study. 
Source: Author.\label{tab1}}
\begin{adjustwidth}{-\extralength}{0cm}
\begin{tabularx}{\fulllength}{llXXl}
\toprule
\textbf{Variable Category} & 
\textbf{Sub-Variable} & 
\textbf{Valence (V)} & 
\textbf{Arousal (A)} & 
\textbf{Publication} \\
\midrule

\multirow{5}{*}{\textbf{Complexity}}
& Fractal Dimension          
  & $\surd$ (V) *** 
  & $\surd$ (A) *** 
  & \citet{Valentine2024} \\
& Silhouette                 
  & $\surd$ (V) *** 
  & $\surd$ (A) *** 
  & \citet{Aydin2022} \\
& Visual complexity          
  & $\surd$ (V) *** 
  & $\surd$ (A) *** 
  & \citet{HashemiKashani2023} \\
& Ornamentation \& Articulation  
  & $\surd$ (V) *** 
  & $\surd$ (A) **  
  & \citet{HashemiKashani2023} \\
& Massing Variation (Form)   
  & $\surd$ (V) **  
  & $\surd$ (A) *** 
  & \citet{Heath2000a} \\

\midrule

\multirow{3}{*}{\textbf{Transparency}}
& Window-to-Wall Ratio       
  & $\surd$ (V) *** 
  & $\surd$ (A) *** 
  & \citet{Hollander2020} \\
& Visual Permeability        
  & $\surd$ (V) **  
  & $\surd$ (A) **  
  & \citet{Akalin2009} \\
& Window Arrangement         
  & --              
  & $\surd$ (A) *   
  & \citet{Lindal2013} \\

\midrule

\multirow{5}{*}{\textbf{Materiality}}
& Surface Roughness          
  & $\surd$ (V) *** 
  & $\surd$ (A) *** 
  & \citet{Stamps2002} \\
& Vegetation on fa\c{c}ade  
  & $\surd$ (V) **  
  & $\surd$ (A) **  
  & \citet{Hollander2020} \\
& Material texture           
  & $\surd$ (V) **  
  & $\surd$ (A) **  
  & \citet{Stamps2002} \\
& Fa\c{c}ade colour         
  & $\surd$ (V) **  
  & $\surd$ (A) **  
  & \citet{Zhu2024} \\
& Colour of tone             
  & $\surd$ (V) *   
  & --              
  & \citet{Shemesh2017} \\

\midrule

\multirow{2}{*}{Environmental Influence}
& Green view factor          
  & $\surd$ (V) *** 
  & $\surd$ (A) *** 
  & \citet{Zhao2024} \\
& Vehicle presence           
  & $\surd$ (V) *** 
  & $\surd$ (A) *** 
  & \citet{Zhao2024} \\

\bottomrule
\end{tabularx}
\end{adjustwidth}
\noindent{\footnotesize{*** $p < .001$; 
** $p < .01$; * $p < .05$; 
-- not significant or not reported.}}
\end{table}

\subsubsection{Complexity}

Fa\c{c}ade complexity, encompassing formal articulation, 
rhythmic subdivision, ornamental density, curvature, and 
material contrast, has been identified as among the most 
consistent perceptual predictors of affective response in 
built environment research. The theoretical basis for this 
relationship derives from Berlyne's psychobiological model 
of aesthetic preference \citep{Berlyne1971}, in which visual 
complexity functions as a collative variable modulating 
arousal potential. Empirical work confirms a robust positive 
association between complexity and subjective arousal across 
visual stimuli \citep{Marin2016}, while the relationship 
between complexity and valence is better characterised as 
curvilinear: stimuli at intermediate levels of visual 
information density tend to yield the highest pleasantness 
ratings, consistent with theories of optimal stimulation 
and processing fluency \citep{Reber2004, Imamoglu2000}.

In architectural contexts specifically, \citet{HashemiKashani2023} 
demonstrate using discrete choice experiments with real 
fa\c{c}ade photographs that features including ornamentation, 
material contrast, and curved geometries significantly enhance 
perceived complexity and positively influence preference. 
Evidence from eye-tracking studies further indicates that 
highly complex fa\c{c}ades demand greater visual effort and 
modulate attentional allocation, suggesting that affective 
response is mediated not only by informational richness but 
by the degree of perceptual organisation available to the 
viewer \citep{Beder2024a}. While vegetation and surface 
greening introduce visual texture that can partially augment 
perceived complexity \citep{White2011}, their affective 
consequences are largely driven by naturalness associations 
rather than by formal compositional properties, and their 
contribution is analytically separable from the geometric 
and morphological dimensions of complexity examined here.

\subsubsection{Transparency}

Transparency, operationalised primarily through window-to-wall 
ratio (WWR) and related measures of visual permeability, 
regulates the perceived degree of connection between interior 
and exterior environments and modulates affective comfort 
through the psychophysiology of spatial openness and prospect. 
Theoretical grounding derives from Gibson's affordance 
framework, in which transparent surfaces signal environmental 
legibility and navigational possibility \citep{Gibson1979}, 
and from prospect-refuge theory, in which visual access to 
the surrounding environment reduces uncertainty and supports 
positive affective states \citep{Kaplan1995}.

Empirically, increased WWR has been associated with reduced 
perceived oppressiveness in high-rise fa\c{c}ade conditions 
\citep{WangMunakata2024}, improved stress recovery 
\citep{Aries2010}, and higher positive affect under controlled 
exposure \citep{Lindal2013}. These effects are consistently mediated by the 
quality of visual access rather than by geometric 
ratio alone: depth cues, layering, and the 
perceptual penetrability of the fenestration 
pattern modulate whether increased transparency 
translates into affective benefit 
\citep{WangMunakata2024, Lindal2013}. This contingency motivates a 
curvilinear expectation for the transparency-arousal 
relationship, where moderate permeability sustains visual 
engagement, alongside a more monotonically positive 
association with valence reflecting the hedonic benefits 
of openness and prospect. Contextual variables such as 
sky exposure fraction and the visual content beyond the 
fa\c{c}ade plane can further modulate these effects 
\citep{WangMunakata2023}, though their contribution 
operates through spatial configuration rather than 
through fa\c{c}ade-level transparency per se, and falls 
outside the analytical scope of the present study.

\subsubsection{Materiality}

Materiality encompasses the tangible surface properties of 
fa\c{c}ade elements, including texture, reflectance, and 
chromatic character, as well as their perceptual and 
associative attributes relating to naturalness, warmth, and 
tactile quality. Its affective relevance is grounded in 
biophilic and stress recovery theories, which propose that 
sensory exposure to natural materials activates 
parasympathetic pathways and attenuates physiological stress 
markers through evolutionary associations with safety and 
resource availability \citep{Ulrich1991, Kaplan1995}.

Empirical evidence increasingly supports a direct role of 
surface materiality in shaping affective outcomes. 
Neurophysiological studies demonstrate that wooden interiors 
and natural surface finishes elicit increased alpha-band 
activity and positive valence ratings relative to artificial 
or synthetic alternatives \citep{Yin2023}. Within architectural contexts, natural surface 
finishes and biophilic interior elements are 
associated with improved psychophysiological 
stress recovery, enhanced positive affect, and 
higher cognitive performance relative to 
synthetic alternatives \citep{Yin2023}, while 
fa\c{c}ade colour and material composition have 
been shown to influence approach-avoidance 
tendencies and aesthetic appraisal \citep{Zhu2024}. These associations motivate a 
positive relationship between the proportion of natural 
surface material and both valence and arousal, with the 
latter expected to plateau at moderate-to-high levels of 
naturalness as novelty effects attenuate and restorative 
effects stabilise \citep{Kaplan1995, Ulrich1991}. Unlike 
complexity and transparency, which engage primarily 
geometric and spatial processing pathways, materiality 
activates both direct sensory and associative evaluative 
routes \citep{Chatterjee2014}, integrating perceptual 
surface cues with culturally and biologically embedded 
meanings. This dual processing pathway may account for 
the consistently strong and contextually stable influence 
of material composition on affective valence observed 
across diverse architectural settings. Surface greening 
and the integration of living vegetation, while sharing 
the naturalness dimension of materiality, introduce 
ecological and spatial variables, including canopy 
density and seasonal variation, that extend beyond 
the static surface properties characterised here 
\citep{White2011, Tost2015}; their affective contributions 
are therefore treated as analytically distinct and are 
not conflated with the materiality construct in the 
present framework.

\section{Methods}
\begin{figure}[H]
\begin{adjustwidth}{-\extralength}{0cm}
\centering
\includegraphics[width=\linewidth]{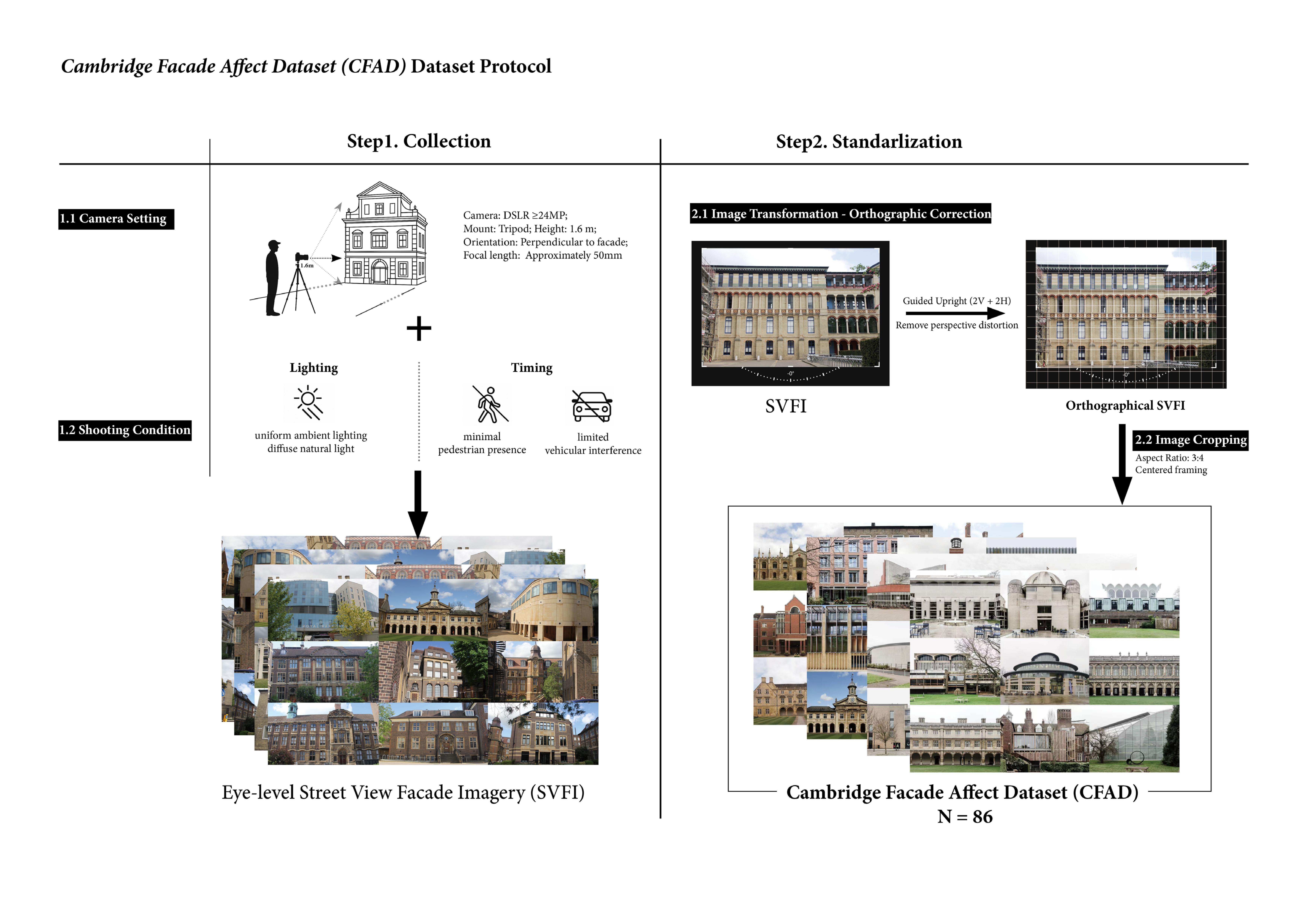}
\caption{Dataset protocol for the Cambridge 
Fa\c{c}ade Affect Dataset (CFAD). Two-step 
protocol for constructing the CFAD image dataset 
($N = 86$), including data collection (Step 1) 
and standardisation (Step 2). Photographs were 
taken using a fixed DSLR setup under uniform 
daylight and minimal urban interference. Images 
underwent orthographic correction and cropping 
to ensure consistent alignment, aspect ratio 
(3:1), and fa\c{c}ade centrality. The resulting 
dataset comprises high-resolution, eye-level, 
rectified street view fa\c{c}ade images for 
experimental use. Source: Author.\label{fig9}}
\end{adjustwidth}
\end{figure}
\unskip
\vspace{12pt}

\begin{figure}[H]
\begin{adjustwidth}{-\extralength}{0cm}
\centering
\includegraphics[width=\linewidth]{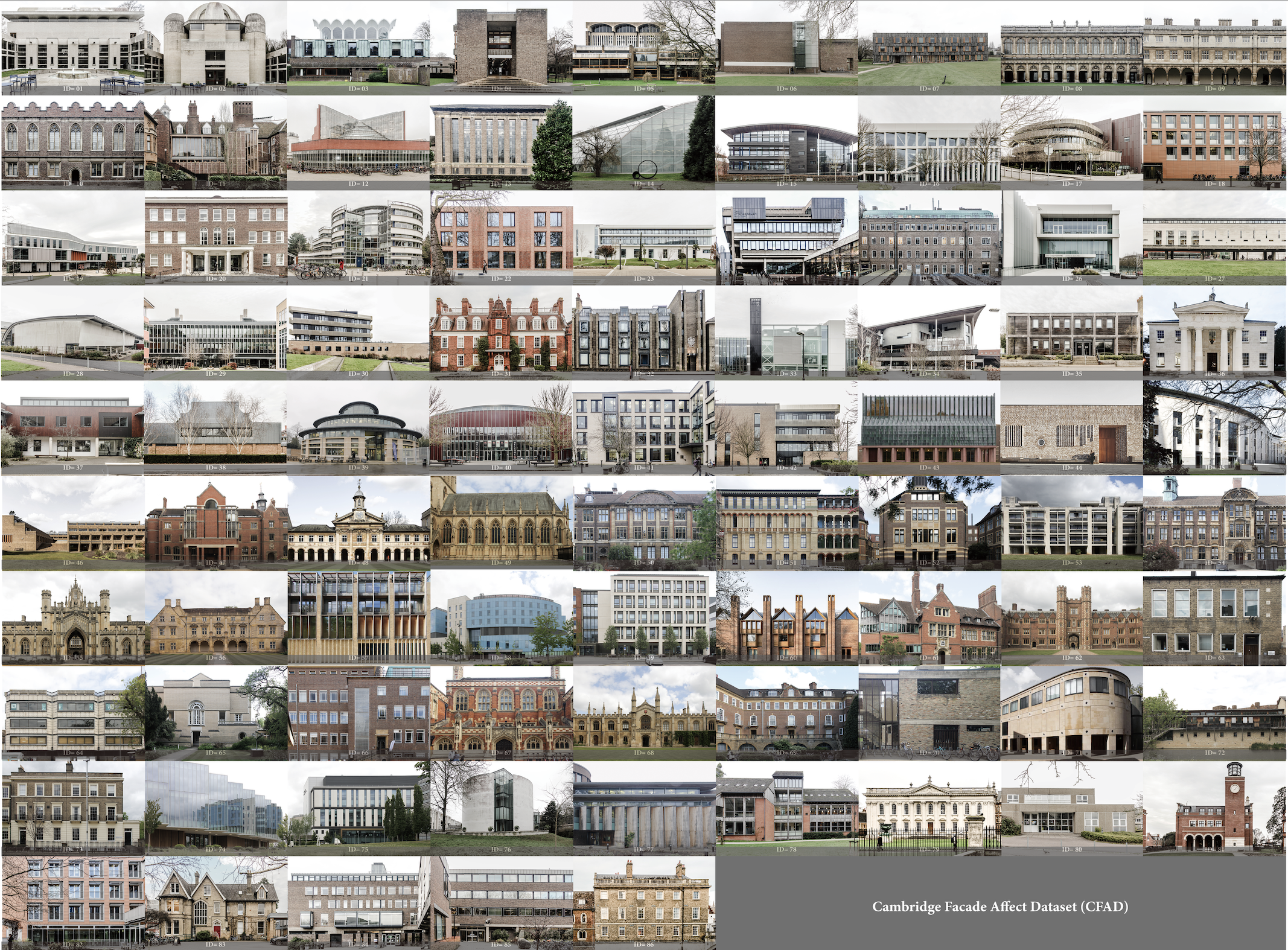}
\caption{Cambridge Fa\c{c}ade Affect Dataset 
(CFAD). Source: Author.\label{fig_cfad}}
\end{adjustwidth}
\end{figure}
\unskip
\vspace{6pt}

\subsection{Computational Measurement of Fa\c{c}ade Attributes}

Three fa\c{c}ade attributes, complexity, transparency, and 
materiality, were extracted computationally from the 86 
orthogonally rectified images comprising the Cambridge 
Fa\c{c}ade Affect Dataset (CFAD). All metrics were normalised 
to a continuous range of $[0, 1]$ to enable systematic 
cross-fa\c{c}ade comparison and integration with perceptual 
ratings. The pipeline was designed to be reproducible and 
scalable, producing attribute scores that serve as the 
primary machine-derived predictors in subsequent 
affective modelling.

\subsubsection{Complexity}

Fa\c{c}ade complexity was operationalised through two 
complementary descriptors capturing visual structure at 
different perceptual scales: edge density and fractal 
dimension.

Edge density was computed by applying Canny edge detection 
to grayscale fa\c{c}ade images using OpenCV, following the 
theoretical grounding that the human visual system is 
selectively sensitive to luminance discontinuities as 
primary carriers of structural information \citep{Marr1980}. 
Edge density was defined as the ratio of detected edge 
pixels to total fa\c{c}ade pixels:

\begin{equation}
C_{\mathrm{edge}} = \frac{N_{\mathrm{edge}}}{N_{\mathrm{total}}}
\end{equation}

This metric captures high-frequency visual transitions 
associated with fine-grained surface articulation and 
local structural richness. To complement this local 
descriptor, fractal dimension $D$ was estimated using 
the box-counting method \citep{Zhao2024}, providing a 
scale-invariant measure of hierarchical complexity and 
nested self-similarity:

\begin{equation}
D = \lim_{\epsilon \to 0} 
    \frac{\log N(\epsilon)}{\log (1/\epsilon)}
\end{equation}

Higher values of $D$ reflect greater multi-scale 
structural organisation, consistent with its established 
application in architectural morphology and urban 
landscape analysis \citep{Stamps2002}. Together, edge 
density and fractal dimension provide a dual-resolution 
representation of fa\c{c}ade complexity, capturing both 
localised visual transitions and global compositional 
hierarchy. Edge density was selected as the primary 
machine-derived complexity score for subsequent 
analysis on the basis of its consistency and 
interpretability across the dataset; fractal dimension 
is retained as a complementary descriptor. Observed 
complexity scores ranged from $0.08$ to $0.72$ 
across the CFAD.

\subsubsection{Transparency}

Transparency was operationalised as the window-to-wall 
ratio (WWR), defined as the proportion of visually 
permeable surface area relative to total fa\c{c}ade 
area:

\begin{equation}
T = \frac{A_{\mathrm{window}}}{A_{\mathrm{facade}}}
\end{equation}

Three segmentation-based window detection approaches 
were evaluated: a YOLO--SAM hybrid combining object 
detection with mask refinement; a YOLOv11 model 
pre-trained on fa\c{c}ade datasets; and a semantic 
segmentation model trained on the COCO fa\c{c}ade 
subset. Each approach reflects a different trade-off 
between detection specificity, generalisation across 
fa\c{c}ade typologies, and sensitivity to occlusion 
and surface reflectance. All outputs were manually 
verified against fa\c{c}ade contours to ensure 
geometric consistency, motivated by evidence that 
WWR estimation is particularly sensitive to fenestration 
diversity and reflective glazing conditions 
\citep{Tarkhan2025}. Among the three approaches, 
the YOLOv11-based pipeline demonstrated the most 
stable performance across the dataset and was 
adopted for final analysis. Transparency scores 
ranged from $0.07$ (near-opaque masonry fa\c{c}ades) 
to $0.62$ (predominantly glazed fa\c{c}ades). 
Implementation details are provided in Appendix~C.

\subsubsection{Materiality}

Materiality was operationalised as the proportion of 
natural relative to artificial surface material, 
reflecting the theoretical distinction between 
biophilically resonant and synthetic surface 
compositions \citep{Ulrich1991, Kaplan1995}. To 
extract material composition without requiring 
manually annotated training data, a zero-shot 
segmentation approach was employed using CLIPSeg, 
a vision--language model that generates class-specific 
attention masks from natural language prompts 
\citep{Luddecke2022}. This approach has been 
validated for large-scale fa\c{c}ade material 
mapping across diverse urban contexts, demonstrating 
robust classification performance across material 
categories including brick, stone, wood, glass, 
tile, and metal \citep{Tarkhan2025}.

Each fa\c{c}ade image was evaluated using six 
predefined material prompts corresponding to these 
categories. A natural material ratio was subsequently 
computed as the proportion of pixels classified as 
natural materials (brick, stone, wood) relative to 
total fa\c{c}ade area. Model outputs were validated 
on a held-out image subset and calibrated to reduce 
sensitivity to lighting variation, surface reflectance, 
and shadow artefacts. Natural material ratios across 
the CFAD ranged from $0.00$ (fully artificial 
fa\c{c}ades) to $0.93$ (predominantly natural 
fa\c{c}ades). 

\begin{figure}[H]
\begin{adjustwidth}{-1cm}{0cm}
\centering
\includegraphics[width=\linewidth]{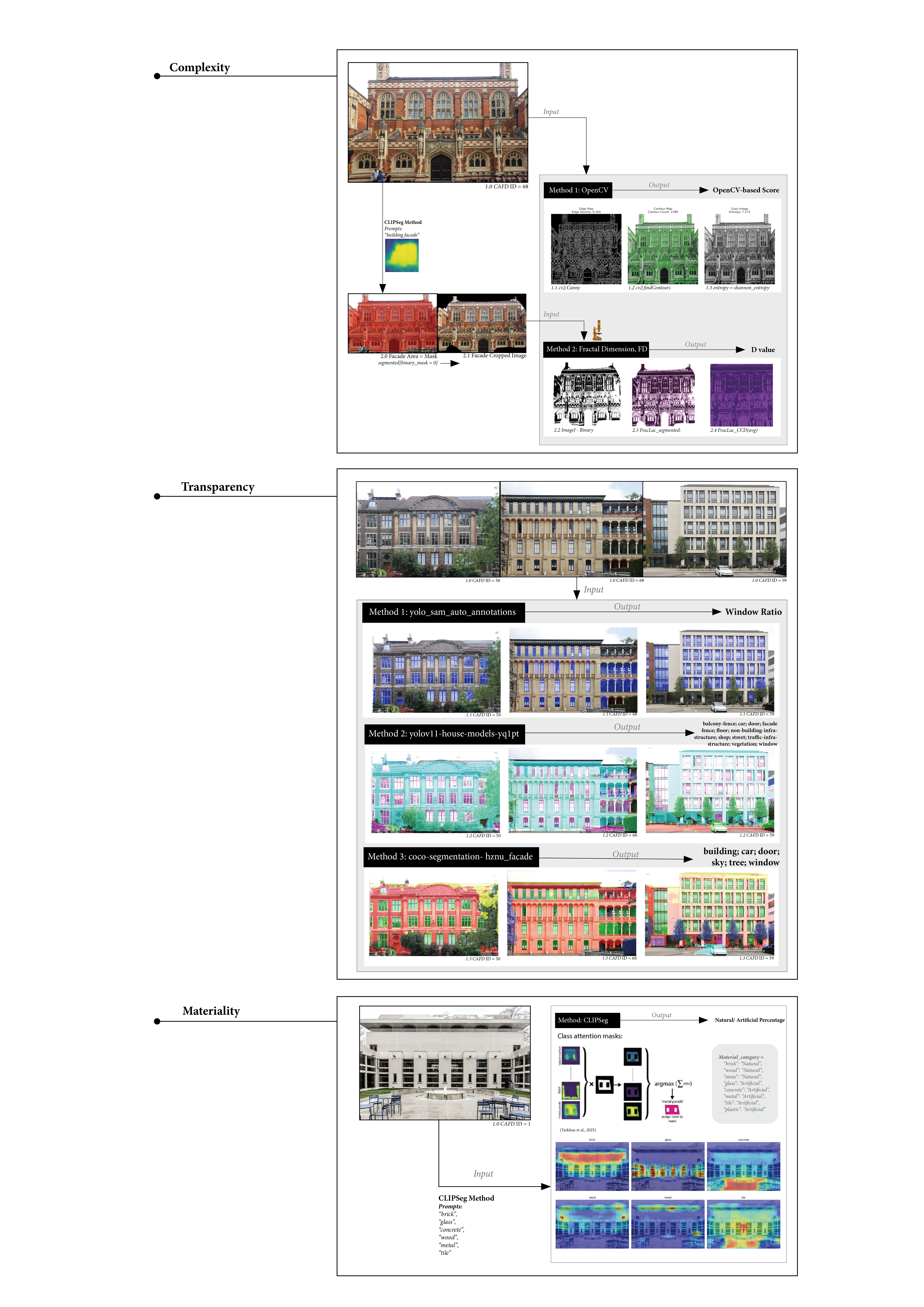}
\caption{Computational quantification of 
fa\c{c}ade features: complexity, transparency, 
and materiality. Three architectural variables 
were quantified from fa\c{c}ade imagery using 
a combination of computer vision techniques. 
Complexity was derived via OpenCV edge density 
and fractal dimension ($D$) methods. 
Transparency was estimated by calculating the 
window-to-wall ratio (WWR) using YOLO and 
semantic segmentation (e.g., COCO, HZNU 
datasets). Materiality was computed as the 
ratio of natural to total surface area using 
CLIPSeg-based zero-shot texture classification. 
Visual examples show input images, algorithmic 
processing steps, and final outputs. 
Source: Author.\label{fig11}}
\end{adjustwidth}
\end{figure}
\unskip
\vspace{6pt}

\begin{figure}[H]
\begin{adjustwidth}{-\extralength}{0cm}
\centering
\includegraphics[width=\linewidth]{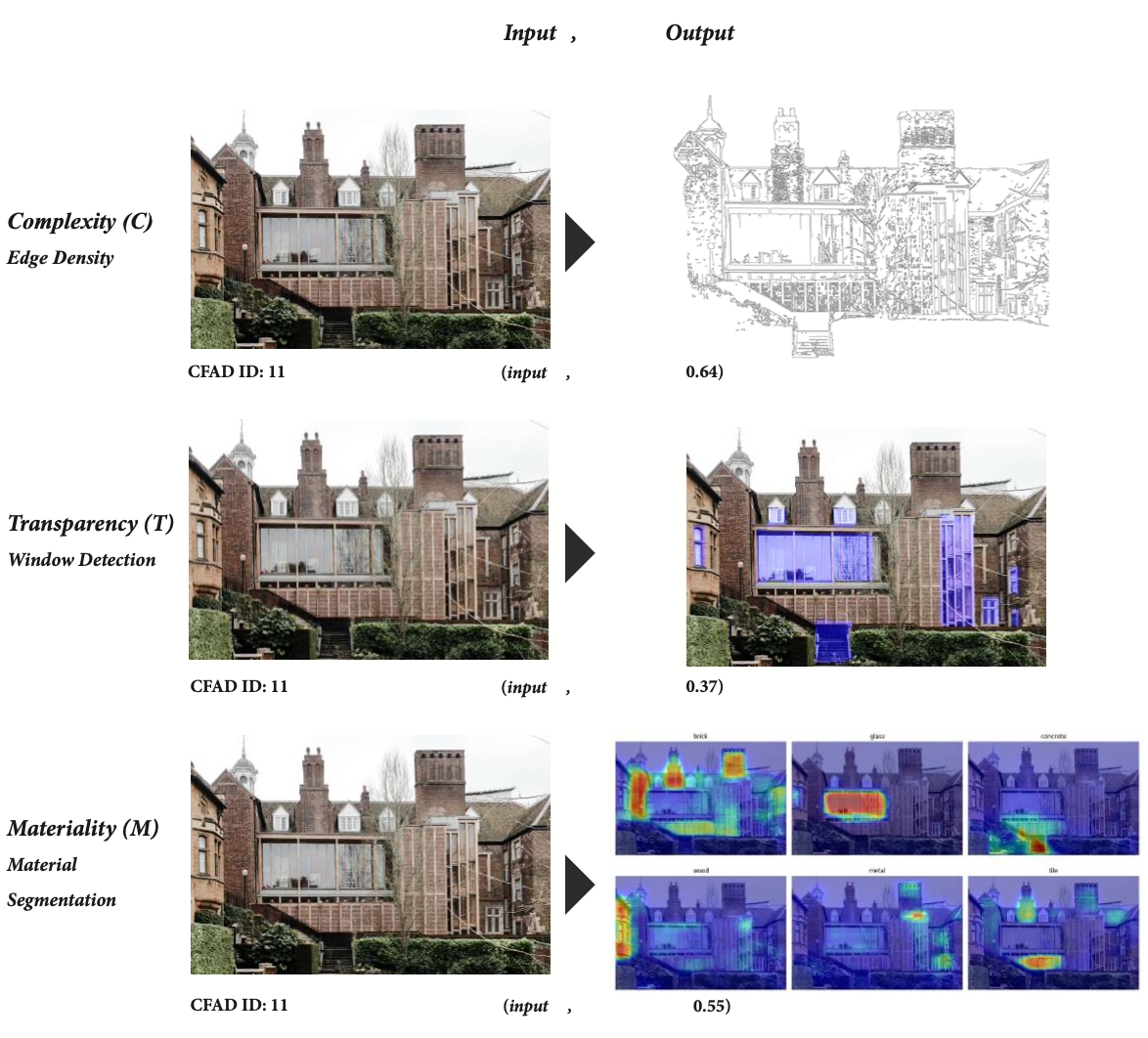}
\caption{Sample pipeline of machine-derived 
fa\c{c}ade feature extraction (CFAD ID: 11). 
Example showcasing automated processing of a 
single fa\c{c}ade image to extract three visual 
metrics. Complexity is computed via edge 
density, transparency via window detection and 
area ratio, and materiality via class-based 
segmentation of natural versus artificial 
materials. Final scores ($0$--$1$) are displayed 
for each variable. This image represents 
fa\c{c}ade ID~11 in the CFAD dataset. 
Source: Author.\label{fig12}}
\end{adjustwidth}
\end{figure}
\unskip
\vspace{6pt}
\subsection{Perceptual Survey: Online and Field-Based Evaluation}

\subsubsection{Survey Design and Stimulus Sampling}
\begin{figure}[H]
\begin{adjustwidth}{-\extralength}{0cm}
\centering
\includegraphics[width=\linewidth]{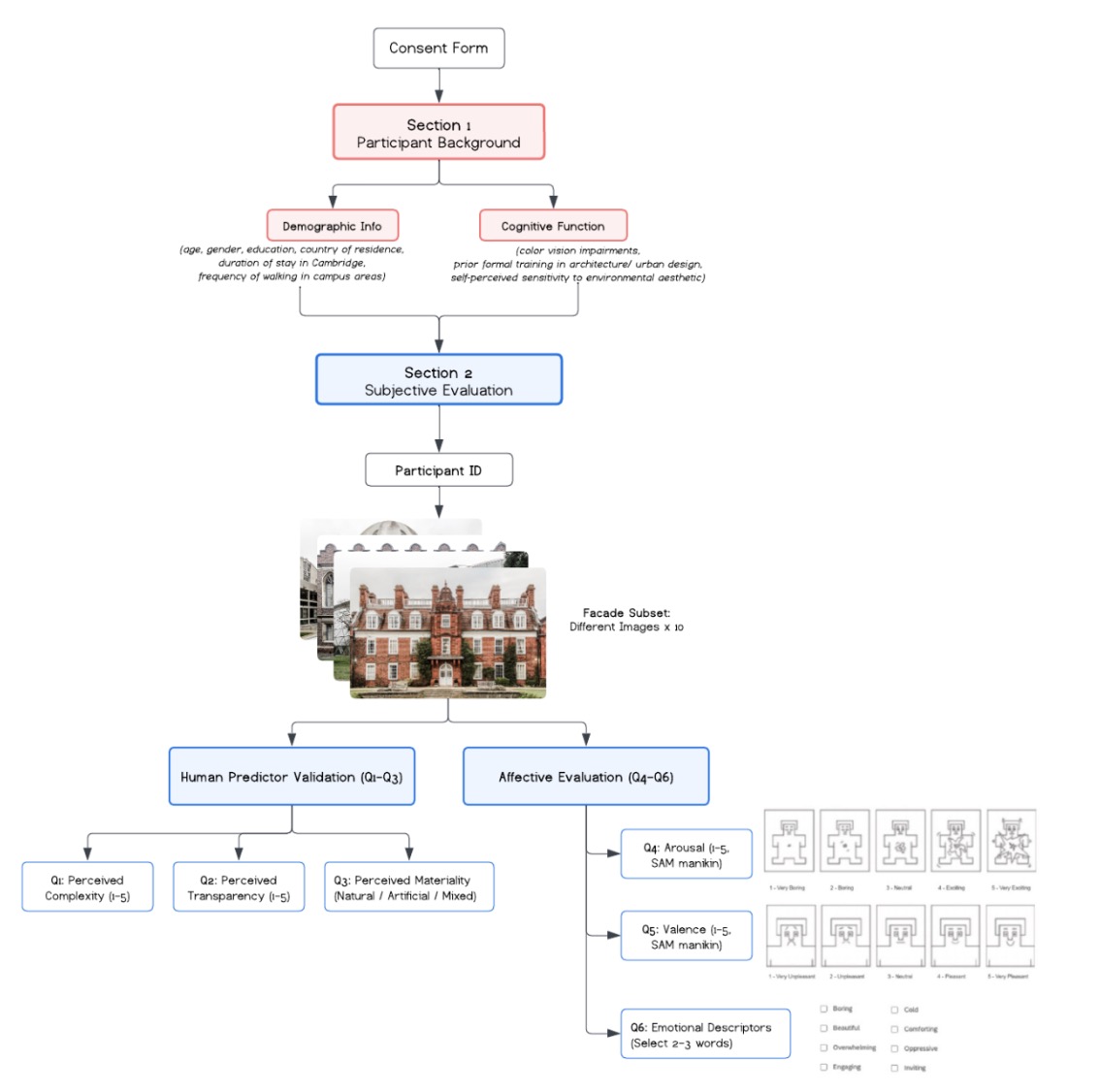}
\caption{Survey workflow for online subjective 
evaluation. Participants ($N_p = 85$) completed 
an online survey comprising two sections: 
background profiling (demographics, cognitive 
traits) and fa\c{c}ade evaluation. Each 
participant was randomly assigned 10 fa\c{c}ade 
images from the CFAD dataset and asked to rate 
them along three perceptual dimensions(complexity, 
transparency, and materiality) as well as two 
affective dimensions (arousal and 
valence) using standardised 5-point Likert 
scales with visual anchors. 
Source: Author.\label{fig14}}
\end{adjustwidth}
\end{figure}
\unskip
\vspace{6pt}
Perceptual and affective evaluations of the 86 CFAD 
fa\c{c}ades were collected through an image-based 
survey administered in two conditions: an online 
platform and a field-based in-situ protocol, enabling 
assessment of both perceptual alignment and ecological 
validity. Given the size of the stimulus set relative 
to the cognitive capacity of individual participants, 
a Balanced Incomplete Block Design (BIBD) was adopted: 
each participant evaluated 15 fa\c{c}ades, while each 
fa\c{c}ade received ratings from at least 12 independent 
participants, yielding 1,275 fa\c{c}ade-level 
observations across the online sample ($N_p = 85$). 
Stimulus assignment was stratified according to 
machine-derived complexity, transparency, and 
materiality scores, grouped into low, medium, and 
high tertiles, ensuring sufficient perceptual variation 
within each participant's allocated set. Presentation 
order was counterbalanced across participants to 
mitigate sequence effects.

Participants first completed a brief demographic 
profile, including age, gender, education level, 
duration of residence in Cambridge, campus visit 
frequency, colour vision status, and self-reported 
architectural background. For each fa\c{c}ade image, 
participants evaluated perceived complexity, 
transparency, and dominant materiality (natural, 
artificial, or mixed). Affective responses were 
subsequently recorded using the Self-Assessment 
Manikin (SAM) scale \citep{Bradley1994}, capturing 
valence (unpleasant--pleasant) and arousal 
(boring--exciting) as continuous nine-point ratings. 
Participants additionally selected two to three 
descriptors from a predefined adjective list to 
enrich affective interpretation and support 
qualitative validation of SAM responses.

\subsubsection{The Self-Assessment Manikin and Affective 
Measurement}

\begin{figure}[H]
\centering
\includegraphics[width=0.5\textwidth]{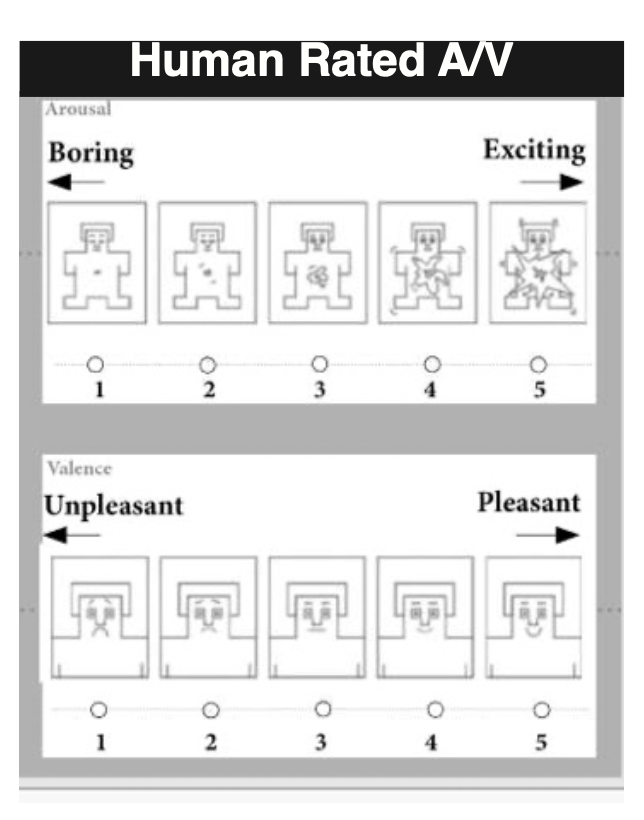}
\caption{Self-Assessment Manikin (SAM) for 
affect measurement. The Self-Assessment Manikin 
(SAM) is a nonverbal pictorial assessment 
technique that directly measures the dimensions 
of affective response. The top row depicts the 
valence scale (from unpleasant to pleasant), 
and the bottom row shows the arousal scale 
(from calm to excited). 
Source: Author.\label{fig16}}
\end{figure}
\unskip
\vspace{6pt}

The SAM scale \citep{Bradley1994} is a non-verbal, 
pictorial instrument that captures affective states 
along the valence and arousal dimensions without 
dependence on linguistic interpretation, making it 
particularly well-suited to cross-context perceptual 
studies and rapid image-based evaluation protocols. 
Its psychometric properties are well-established 
across environmental psychology and neuroarchitecture 
research \citep{Coburn2020, Higuera2022}. The 
non-verbal format aligns with evidence that affective 
reactions to visual stimuli are frequently 
pre-reflective and arise prior to conscious linguistic 
appraisal \citep{Zajonc1984, Pessoa2008}, capturing 
the immediate evaluative layer of environmental 
perception discussed in Section~2.

SAM ratings have been shown to produce reliable 
dimensional affective profiles in response to 
architectural stimuli: parametric manipulation of 
spatial dimensions in virtual environments generates 
distinct, replicable patterns within the 
valence--arousal space \citep{Chiamulera2022}, while 
EEG studies confirm that self-reported arousal and 
valence correspond to separable patterns of cortical 
activation in response to built environment stimuli 
\citep{Banaei2017}. These convergent findings support 
the use of SAM as a validated operational measure of 
dimensional affect in image-based fa\c{c}ade 
evaluation, and its continuous output enables direct 
integration with computationally derived fa\c{c}ade 
metrics in subsequent modelling.

\subsection{Statistical Power and Sample Adequacy}

The BIBD structure was selected as particularly 
appropriate for perceptual studies in which the 
number of stimuli substantially exceeds the number 
of participants, and in which the primary inferential 
focus lies on stimulus-level rather than 
individual-level effects \citep{Westfall2014}. 
Power analysis was conducted using a medium effect 
size benchmark ($f^2 = 0.06$), consistent with 
effect sizes reported in prior fa\c{c}ade-affect 
research \citep{Heath2000a, Akalin2009}. 
Simulation-based studies of partially crossed 
linear mixed-effects models indicate that a minimum 
of 12 observations per stimulus, combined with 
80--100 participants, is generally sufficient to 
achieve statistical power above 0.80 for 
medium-sized effects \citep{Westfall2014, 
Brysbaert2018}. The present design meets these 
criteria, with each of the 86 fa\c{c}ades receiving 
a minimum of 12 independent ratings and 85 
participants contributing a total of 1,275 
observations.

\subsection{Ecological Validity: Field-Based Survey}

To assess the stability of affective judgements 
under real-world exposure conditions, a field-based 
survey was administered to a subset of participants 
($N_p = 19$) who evaluated the same fa\c{c}ades 
in situ at their corresponding locations within 
Cambridge. This protocol enabled direct comparison 
between online image-based ratings and on-site 
evaluations, testing whether affective responses 
derived from orthogonally rectified photographs 
generalise to naturalistic viewing conditions. 
Methodological precedents support such cross-modal 
comparisons: psychophysiological responses to 
photographic and in-situ urban stimuli have been 
shown to correlate significantly when stimulus 
content is controlled \citep{Aspinall2013}, and 
image-based surveys of architectural preference 
demonstrate comparable rating structures to 
direct environmental exposure \citep{Stamps2002}. 
Differences in affective ratings between conditions 
are modelled as a secondary outcome, with 
particular attention to any systematic shifts in 
arousal attributable to multisensory context under 
field conditions.

\section{Results}
\begin{figure}[H]
\begin{adjustwidth}{-\extralength}{0cm}
\centering
\includegraphics[width=\linewidth]{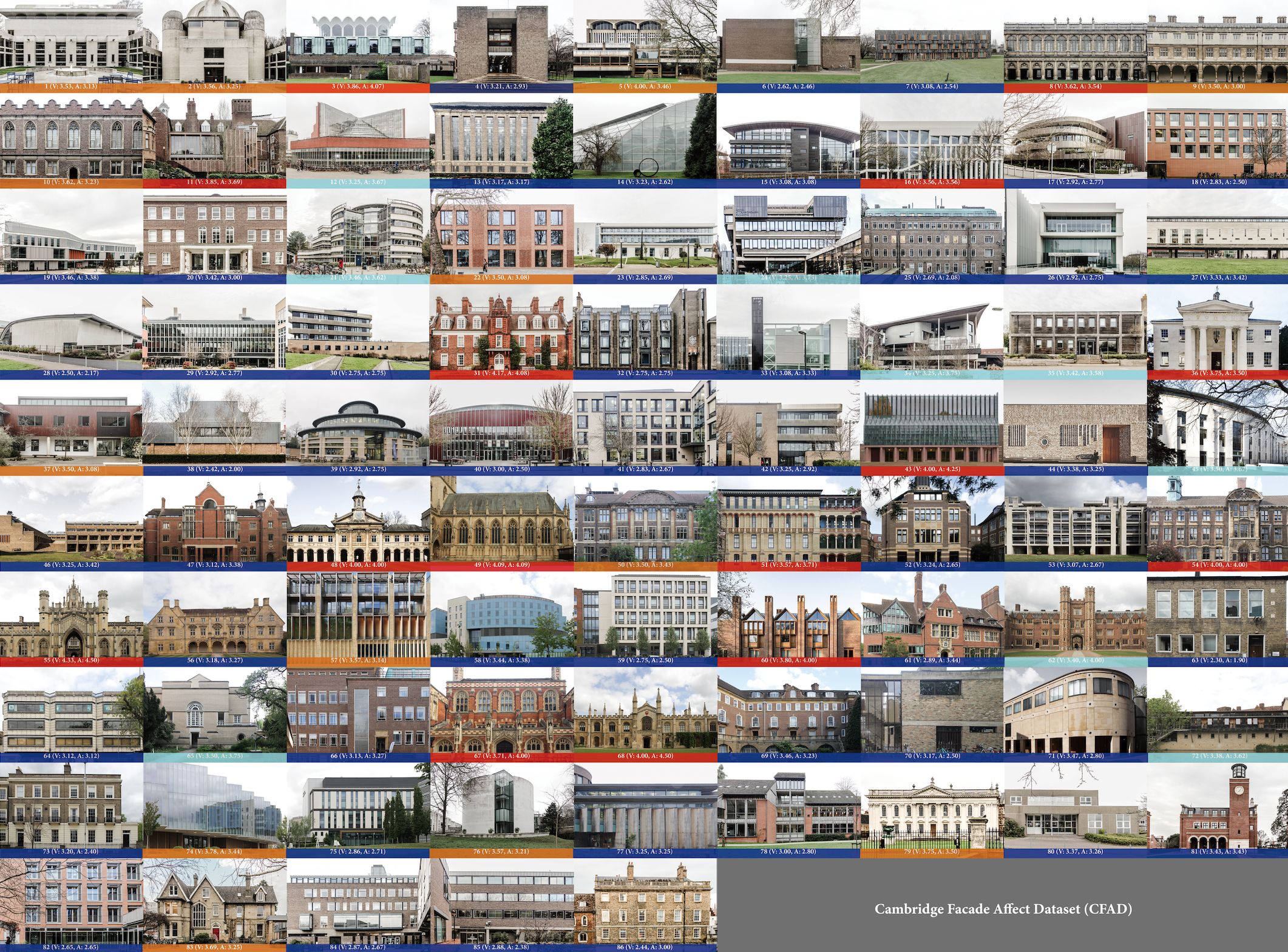}
\caption{ Four emotional 
categories are defined by the combination of 
valence and arousal dimensions: high valence 
and high arousal (pleasant and activating); 
high valence but low arousal (pleasant and 
calm); low valence and low arousal (unpleasant 
and calm); low valence but high arousal 
(unpleasant and activating). Colour coding 
corresponds to affective tone used throughout 
the figures. Source: Author.\label{fig_legend}}
\end{adjustwidth}
\end{figure}
\unskip
\vspace{6pt}

\begin{figure}[H]
\begin{adjustwidth}{-\extralength}{0cm}
\centering
\includegraphics[width=\linewidth]{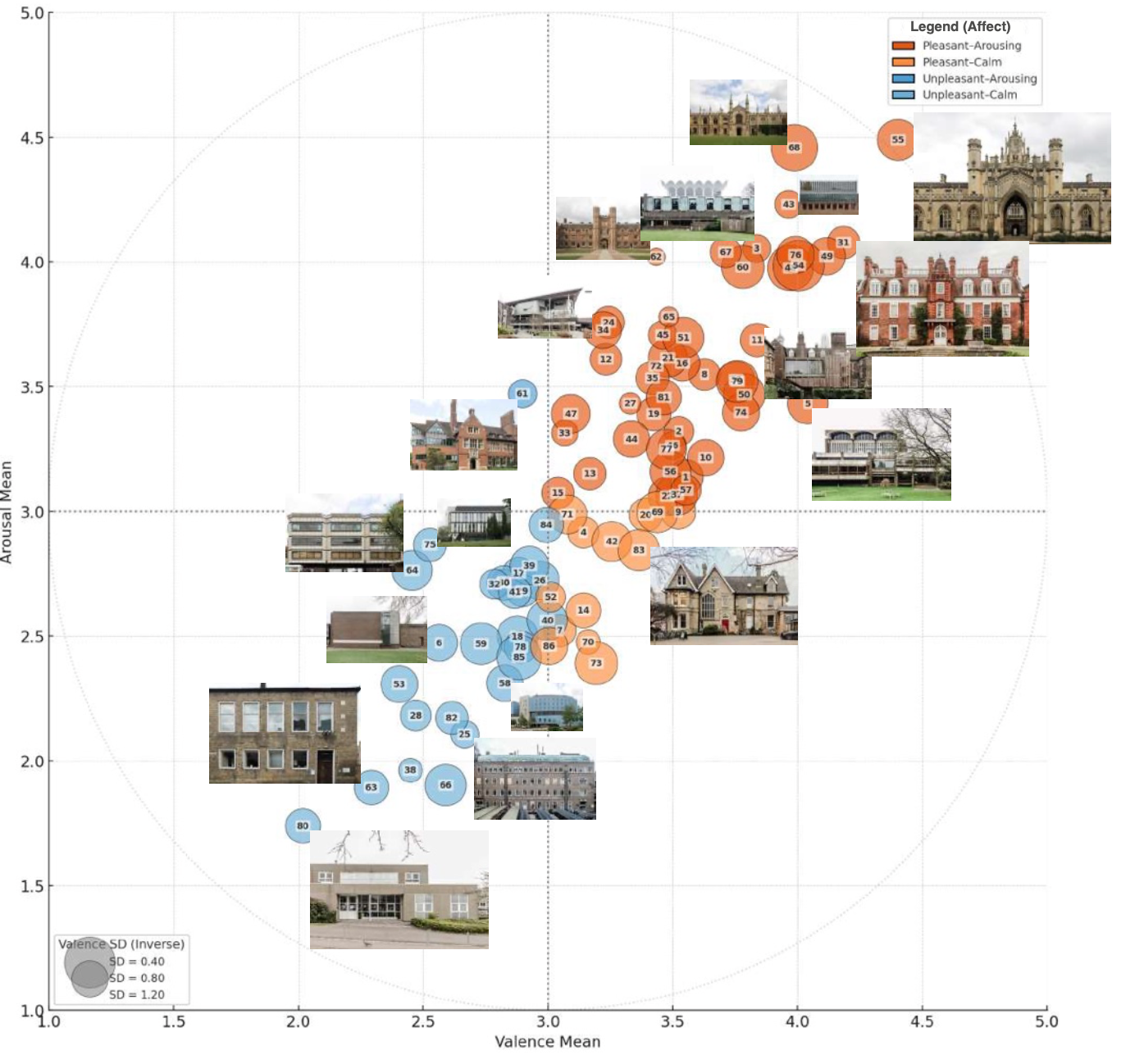}
\caption{Mapping of architectural fa\c{c}ades 
in a two-dimensional affective space. Each 
circle represents one CFAD fa\c{c}ade image 
($N = 86$), positioned by its mean valence 
($x$-axis) and arousal ($y$-axis) ratings 
collected from 85 participants on a 5-point 
Likert scale (1 = low, 5 = high). The space 
is divided into four quadrants reflecting 
classic emotional categories. Circle colour 
indicates the quadrant's affective tone: warm 
colours for pleasant emotions, cool for 
unpleasant. Circle size is inversely scaled 
to the standard deviation (SD) of valence 
ratings --- larger circles reflect stronger 
agreement across participants. The figure 
reveals patterns in collective emotional 
response to built environments. 
Source: Author.\label{fig17}}
\end{adjustwidth}
\end{figure}
\unskip
\vspace{6pt}

\subsection{Descriptive Statistics and Affective 
Distribution}

\begin{table}[H]
\caption{Affective extremes: images with highest 
and lowest valence and arousal ratings. Summary 
of fa\c{c}ade images with the most extreme 
affective responses ($N_f = 86$). Metrics include 
mean valence and arousal scores as well as 
corresponding standard deviations (SD), based on 
online human ratings ($N_p = 85$).\label{tab3}}
\begin{tabularx}{\textwidth}{lCCCCC}
\toprule
& \textbf{Image ID} 
& \textbf{Valence Mean} 
& \textbf{Valence SD} 
& \textbf{Arousal Mean} 
& \textbf{Arousal SD} \\
\midrule
\multirow{2}{*}{\textbf{Highest Valence}} 
& 55 & 4.33 & 0.52 & 4.50 & 0.55 \\
& 31 & 4.17 & 0.90 & 4.08 & 0.76 \\
\midrule
\multirow{2}{*}{\textbf{Lowest Valence}}  
& 80 & 2.00 & 0.82 & 1.75 & 0.96 \\
& 63 & 2.30 & 0.82 & 1.90 & 0.99 \\
\midrule
\textbf{Highest Arousal} 
& 55 & 4.33 & 0.52 & 4.50 & 0.55 \\
\midrule
\textbf{Lowest Arousal}  
& 80 & 2.00 & 0.82 & 1.75 & 0.96 \\
\bottomrule
\end{tabularx}
\noindent{\footnotesize{SD: standard deviation; 
$N_f$: number of fa\c{c}ade images; 
$N_p$: number of participants.}}
\end{table}

Across the 86 CFAD fa\c{c}ades, mean valence scores 
ranged from 2.00 to 4.33 ($M = 3.21$, $SD = 0.76$) 
and arousal scores from 1.75 to 4.50 ($M = 2.95$, 
$SD = 0.72$), indicating substantial stimulus-level 
variability in affective response. At the positive 
extreme, Image~55 (St John's College New Court) and 
Image~31 (Sidgwick Hall, Newnham College) received 
the highest valence ratings; at the negative extreme, 
Image~80 (Kapitza Building, Department of Physics) 
and Image~63 (Faculty of Architecture and History 
of Art, Mill Lane) consistently occupied the lowest 
valence and arousal positions.

When mapped onto the two-dimensional valence--arousal 
space, fa\c{c}ade stimuli exhibit a structured, 
non-random distribution characterised by a diagonal 
clustering pattern, with observations concentrated 
in the pleasant--arousing and unpleasant--calm 
quadrants. Visually articulated fa\c{c}ades with 
high surface variation tended to elicit both higher 
valence and higher arousal, while monotonous or 
visually enclosed fa\c{c}ades produced responses 
in the low-valence, low-arousal region. 
Inter-rater agreement was inversely related to 
affective ambiguity: stimuli at affective extremes 
showed lower standard deviations and stronger 
consensus, whereas centrally located stimuli 
exhibited greater dispersion, reflecting higher 
interpretive variability.

\subsection{Perceptual Attributes and Affective 
Response}
\begin{figure}[H]
\begin{adjustwidth}{-\extralength}{0cm}
\centering
\includegraphics[width=\linewidth]{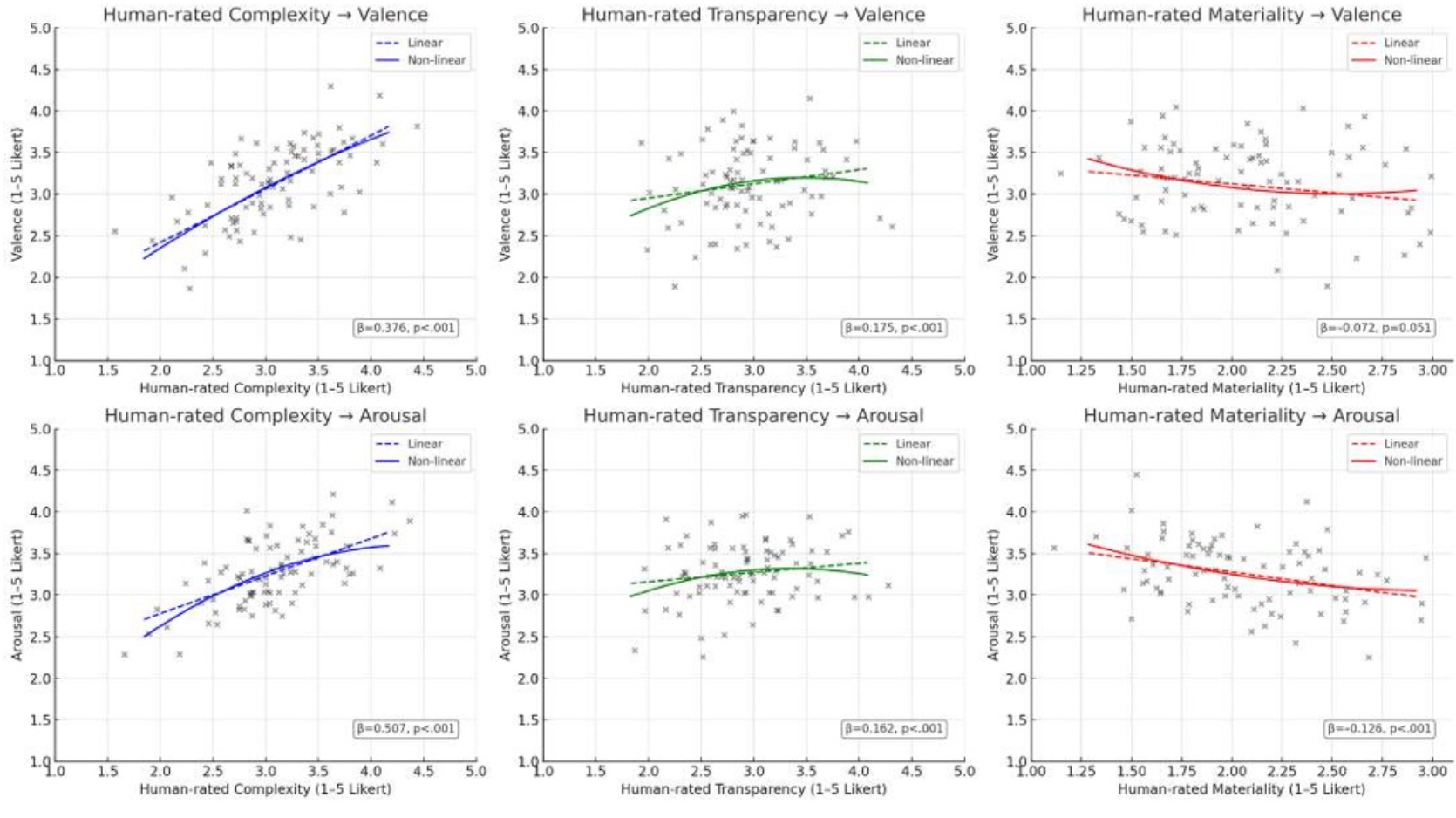}
\caption{Mixed-effects regression models and 
non-linear effects linking perceived fa\c{c}ade 
features to affective responses. Scatterplots 
show the relationship between three human-rated 
fa\c{c}ade attributes---complexity (blue), 
transparency (green), and materiality 
(red)---and affective responses: valence (top 
row) and arousal (bottom row). Data are drawn 
from the CFAD ($N = 86$ fa\c{c}ades), each 
rated by ${\geq}12$ participants ($N_p = 85$) 
using a 5-point Likert scale. Jittered points 
represent individual fa\c{c}ade-level ratings. 
Linear effects were estimated using linear 
mixed-effects models (random intercept for 
participant). Significant fixed effects were 
found for complexity (valence: $\beta = 0.376$, 
$p < 0.001$; arousal: $\beta = 0.507$, 
$p < 0.001$), transparency (valence: 
$\beta = 0.175$, $p < 0.001$; arousal: 
$\beta = 0.162$, $p < 0.001$), and materiality 
(arousal: $\beta = -0.126$, $p < 0.001$). 
Non-linear patterns were modelled with 
second-order polynomial fits (solid lines). 
Complexity showed upward-curving trends; 
transparency revealed an inverted-U relationship 
with valence; materiality exhibited 
concave-negative effects, most pronounced for 
arousal. Source: Author.\label{fig19}}
\end{adjustwidth}
\end{figure}
\unskip
\vspace{6pt}
To examine how human-rated fa\c{c}ade attributes 
relate to affective outcomes, linear mixed-effects 
models (LMEs) with restricted maximum likelihood 
estimation were fitted to the image-level dataset 
($N_p = 85$, $N_f = 86$), with participant 
identity modelled as a random intercept to account 
for between-subject variance.

Perceived complexity was the strongest and most 
consistent predictor across both affective 
dimensions, with significant positive associations 
for arousal ($\beta = 0.507$, $p < 0.001$) and 
valence ($\beta = 0.376$, $p < 0.001$), indicating 
that fa\c{c}ades rated as more complex were 
simultaneously more stimulating and more positively 
evaluated. Polynomial models further revealed a 
non-linear amplification in the 
complexity--valence relationship, with affective 
responses increasing more steeply at higher levels 
of perceived complexity, and without evidence of 
saturation within the observed range. This 
pattern is broadly consistent with Berlyne's 
psychobiological framework \citep{Berlyne1971}, 
though it suggests that the upper limit of the 
inverted-U may lie beyond the complexity levels 
represented in this architectural sample.

Perceived transparency produced weak linear 
effects (arousal: $\beta = 0.088$, $p = 0.420$; 
valence: $\beta = 0.161$, $p = 0.140$), yet 
second-order polynomial models revealed a 
pronounced inverted-U relationship with valence, 
whereby fa\c{c}ades with moderate transparency 
elicited the highest positive affect. This 
curvilinear structure suggests that transparency 
influences affect through perceptual thresholds 
rather than linear accumulation, consistent with 
prospect-refuge accounts of visual access 
\citep{Kaplan1995} in which moderate openness 
is affectively preferred over both occlusion 
and full visual exposure.

Perceived materiality showed a significant 
negative association with arousal ($\beta = -0.126$, 
$p < 0.001$) and a marginal effect on valence 
($\beta = -0.072$, $p = 0.051$), indicating that 
fa\c{c}ades rated as more artificial are associated 
with lower physiological engagement. Polynomial 
models further identified a concave non-linear 
pattern, with steeper affective decline at high 
levels of artificiality, consistent with 
biophilic and stress recovery theories in which 
natural surface compositions facilitate more 
positive and activating perceptual states 
\citep{Ulrich1991, Kaplan1995}.

\begin{figure}[H]
\centering
\includegraphics[width=\textwidth]{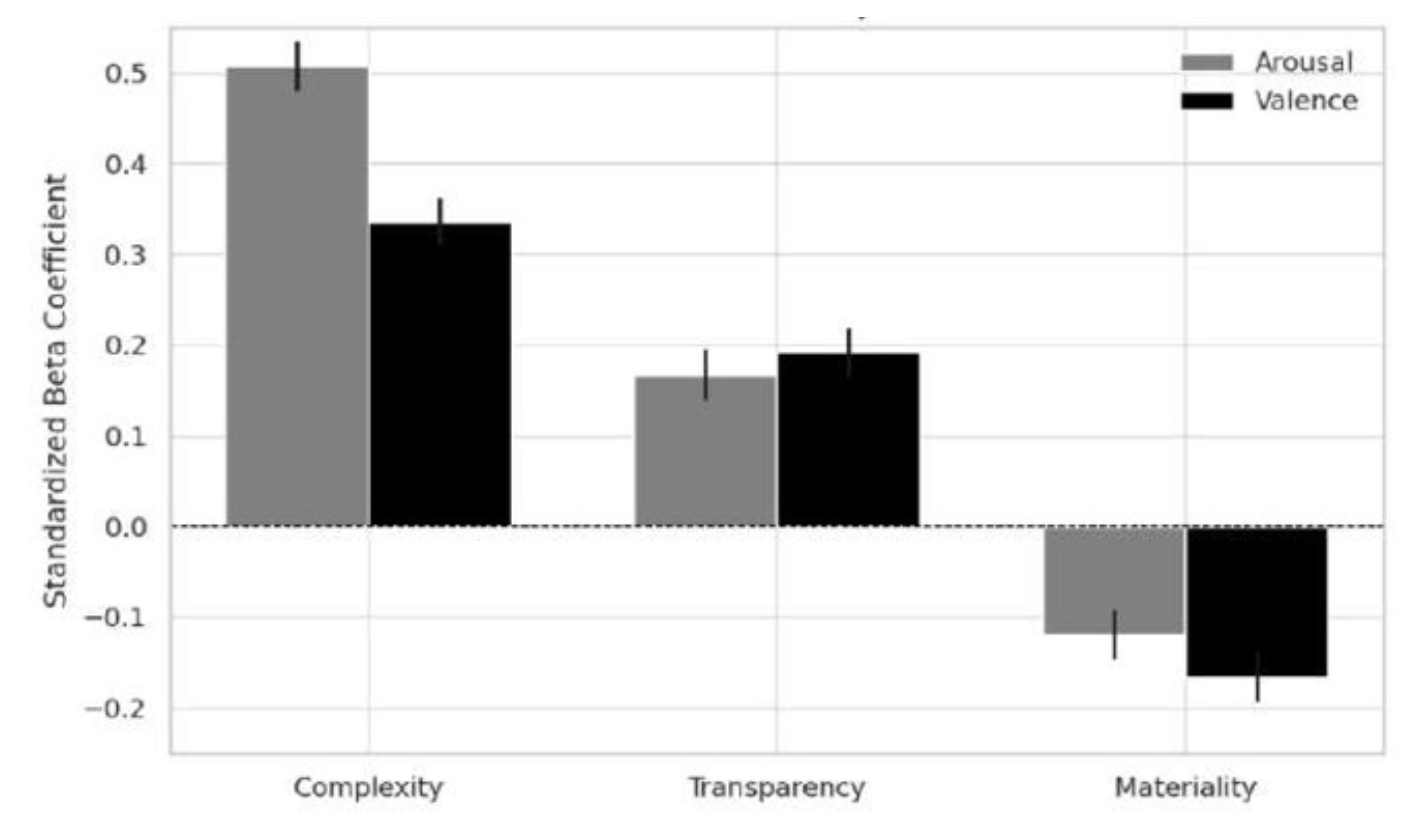}
\caption{Standardised regression coefficients 
for the effects of perceived fa\c{c}ade features 
on arousal and valence. Complexity emerged as 
the strongest positive predictor for both 
affective dimensions; transparency showed weaker 
yet significant positive effects; materiality 
had negative coefficients for both arousal and 
valence, indicating that more natural materials 
(lower materiality score) were associated with 
greater emotional activation and pleasantness. 
Source: Author.\label{fig20}}
\end{figure}
\unskip
\vspace{6pt}

\begin{figure}[H]
\begin{adjustwidth}{-\extralength}{0cm}
\centering
\includegraphics[width=\linewidth]{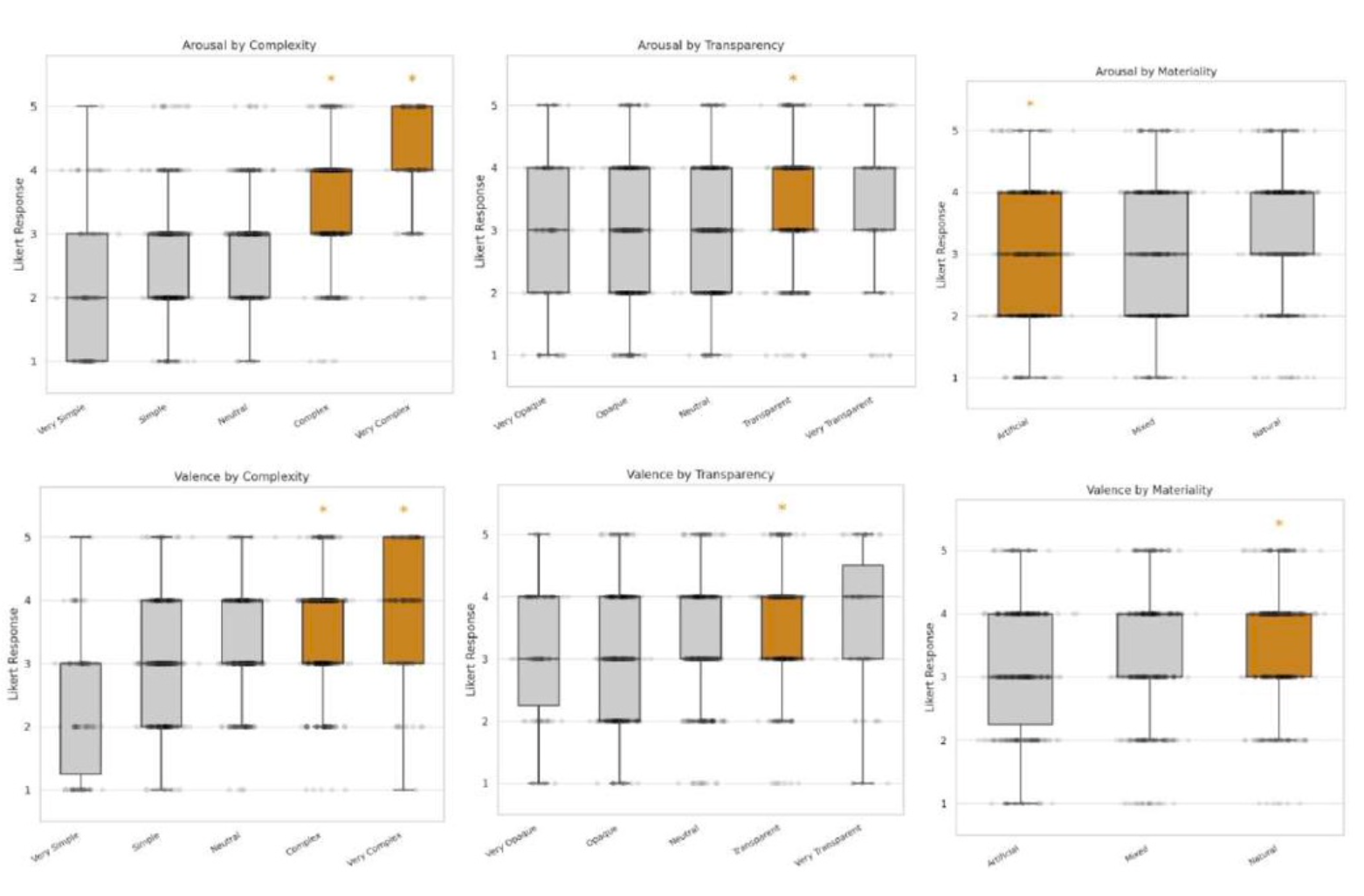}
\caption{Boxplots showing the effects of 
perceived complexity, transparency, and 
materiality on participants' affective 
responses: (\textbf{a}) arousal (excitement) 
and (\textbf{b}) valence (pleasantness). Each 
dot represents an individual Likert rating 
(1--5), jittered for clarity. Boxes indicate 
interquartile ranges (IQR) with medians. Only 
groups with significantly higher median scores 
compared to others (via Kruskal--Wallis test 
and Bonferroni-corrected Mann--Whitney U tests, 
$p < .05$) are highlighted in orange. Higher 
perceived complexity, transparency, and natural 
materiality are associated with increased 
valence (pleasantness), while higher complexity 
and artificial materiality elicit higher arousal 
(excitement). Source: Author.\label{fig21}}
\end{adjustwidth}
\end{figure}
\unskip
\vspace{6pt}

\subsection{Machine-Derived Attributes, Perceptual 
Mediation, and Affective Response}

\begin{figure}[H]
\begin{adjustwidth}{-\extralength}{0cm}
\centering
\includegraphics[width=\linewidth]{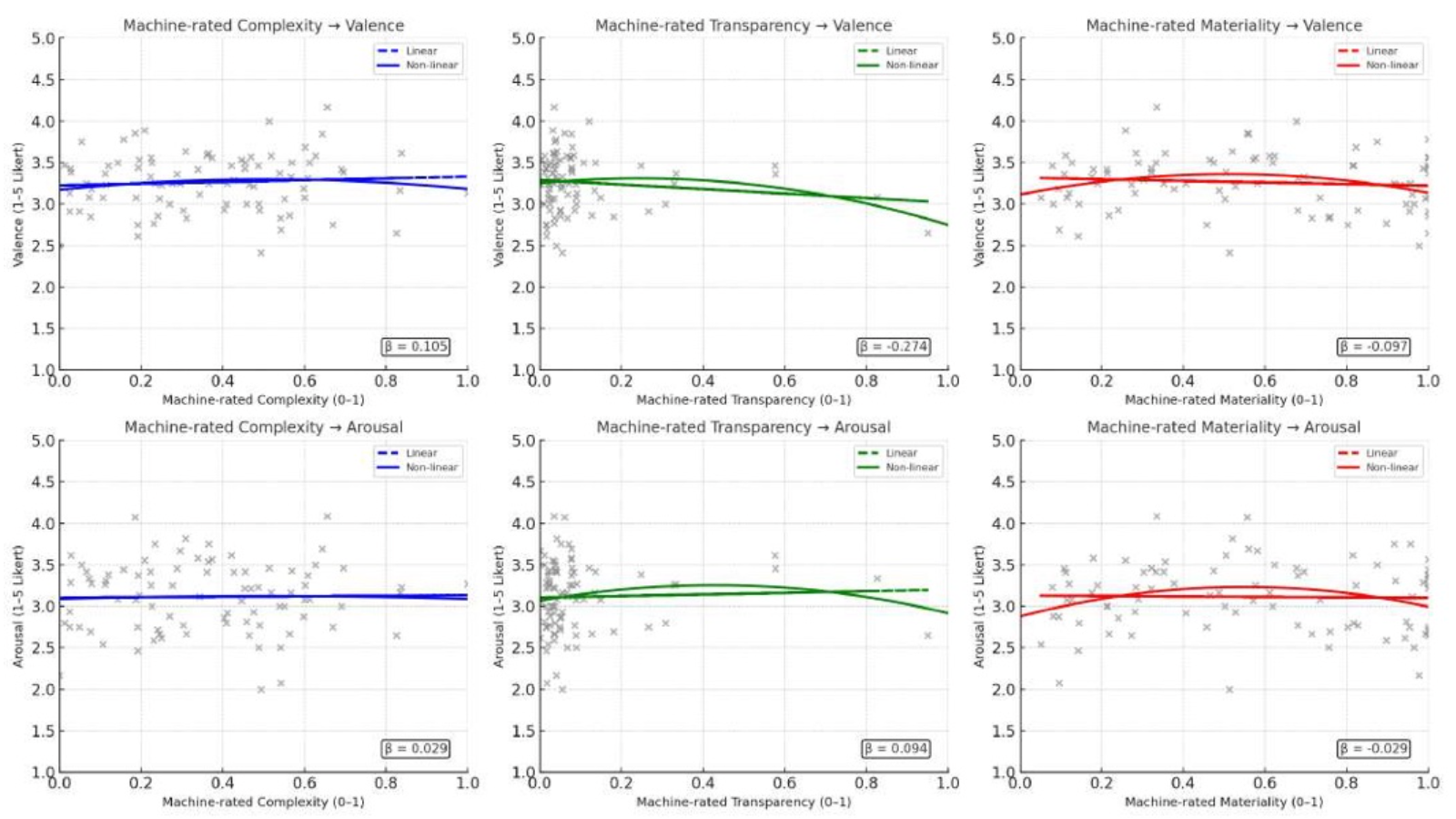}
\caption{Scatter plots showing the relationship 
between machine-derived fa\c{c}ade features and 
affective responses. Each subplot illustrates 
the predictive relationship between a 
machine-derived fa\c{c}ade feature---complexity, 
transparency, or materiality---and one of two 
affective dimensions: valence (top row) or 
arousal (bottom row). Machine ratings were 
computed via computer vision pipelines and 
normalised between 0 and 1. Affective responses 
were collected through online Likert-scale 
surveys (1--5). Dashed lines indicate linear 
regression fits; solid lines represent 
second-degree polynomial (non-linear) models. 
Standardised linear regression coefficients 
($\beta$) are displayed in the lower-right 
corner of each panel. 
Source: Author.\label{fig22}}
\end{adjustwidth}
\end{figure}
\unskip
\vspace{6pt}

Spearman correlation analyses between 
machine-derived fa\c{c}ade attributes and 
image-level mean affective ratings revealed 
limited direct associations. Across five of 
six feature--response pairs, correlations were 
small and non-significant ($p > 0.47$). The 
sole exception was a weak positive association 
between machine-derived materiality and valence 
($\rho = 0.22$, $p = 0.046$). Multivariate 
linear models confirmed this pattern: 
machine-derived features accounted for a 
negligible proportion of variance in both 
valence ($R^2 = 0.024$, $p = 0.579$) and 
arousal ($R^2 = 0.002$, $p = 0.981$).

Mediation analyses revealed that human 
perceptual evaluations function as a 
significant intermediate layer between 
physical fa\c{c}ade attributes and affective 
outcomes. Human-rated materiality significantly 
mediated the machine--valence relationship 
(indirect effect $= -0.205$, $p = 0.003$), 
indicating that the affective impact of 
material composition is carried through its 
perceived naturalness rather than through 
direct stimulus encoding. Human-rated 
complexity exhibited marginal mediation 
effects on both valence ($p = 0.094$) and 
arousal ($p = 0.090$), while transparency 
showed no reliable mediation pathway. These 
findings position perceptual processing as 
a necessary intermediary in the translation 
of physical fa\c{c}ade structure into 
affective response, with direct implications 
for the design of computational models of 
urban affect.

\subsection{Validation}

Three validation analyses examined the 
cross-modal robustness of the measurement 
framework, targeting alignment between 
computational and perceptual representations, 
perceptual stability across viewing 
conditions, and affective consistency 
between online and field-based settings.

\begin{figure}[H]
\centering
\includegraphics[width=\textwidth]{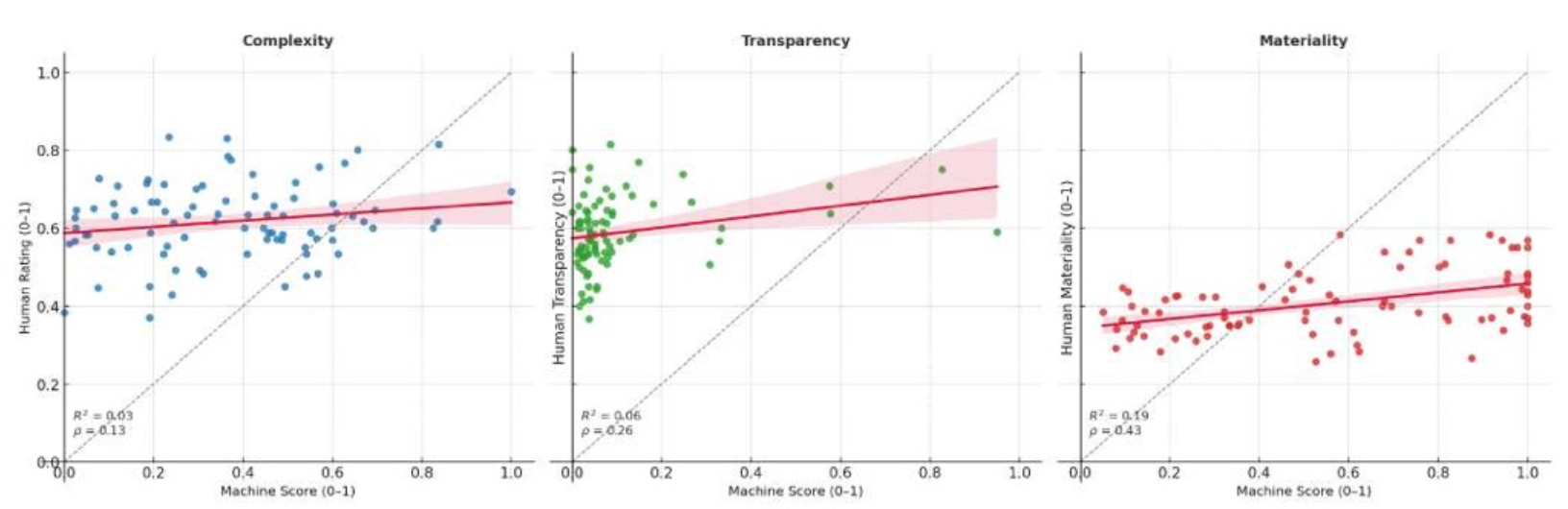}
\caption{Machine--human agreement across three 
fa\c{c}ade dimensions. Scatter plots show the 
correlation between machine-derived and 
human-perceived scores for (\textbf{a}) 
complexity, (\textbf{b}) transparency, and 
(\textbf{c}) materiality, all normalised to a 
$0$--$1$ scale. Red lines represent linear 
fits; grey dashed lines indicate identity lines. 
$R^2$ and Spearman's $\rho$ are shown for each 
dimension. Source: Author.\label{fig24}}
\end{figure}
\unskip
\vspace{6pt}

\begin{figure}[H]
\centering
\includegraphics[width=\textwidth]{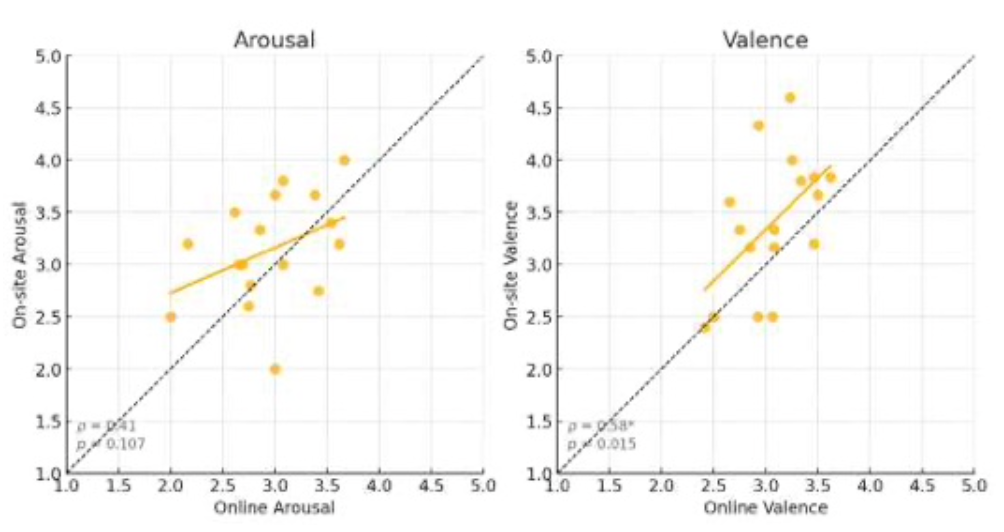}
\caption{Correlation between online and on-site 
affective ratings. Scatter plots with regression 
lines and identity lines (dashed) illustrate 
consistency levels. Valence scores show 
significant correlation ($\rho = 0.58$, 
$p = 0.015$), while arousal scores exhibit 
weaker alignment ($\rho = 0.41$, 
$p = 0.107$). 
Source: Author.\label{fig26a}}
\end{figure}
\unskip
\vspace{6pt}

\textit{Computational--perceptual alignment.} 
Among the three fa\c{c}ade attributes, 
materiality showed the strongest agreement 
between machine-derived and human-rated 
scores ($\rho = 0.431$, $p < 0.001$; 
$r = 0.441$, $p < 0.001$), indicating 
robust cross-modal correspondence for 
surface material classification. Transparency 
showed weaker but statistically significant 
alignment ($\rho = 0.256$, $p = 0.017$), 
while complexity did not reach significance 
($\rho = 0.127$, $p = 0.243$). The 
divergence for complexity suggests that 
human-perceived complexity recruits 
higher-order structural and semantic 
information, including hierarchical 
organisation and perceptual grouping, 
that is not adequately represented by 
low-level edge-based descriptors 
\citep{Doersch2012, Marr1980}.

\textit{Cross-context perceptual stability.} 
In the field-based subsample ($N_f = 15$, 
$N_p = 19$), materiality ratings demonstrated 
the highest cross-context reliability 
(ICC$(2,1) = 0.677$, $p > 0.6$ for paired 
differences), followed by complexity 
(ICC $= 0.556$, $p > 0.8$). Transparency 
ratings exhibited high variability and an 
unstable distributional structure across 
viewing conditions, suggesting that 
transparency perception is sensitive to 
dynamic, embodied cues, including light 
transmission, surface reflectance, and 
spatial depth, that are only partially 
represented in static orthographic images.

\begin{figure}[H]
\centering
\includegraphics[width=\textwidth]{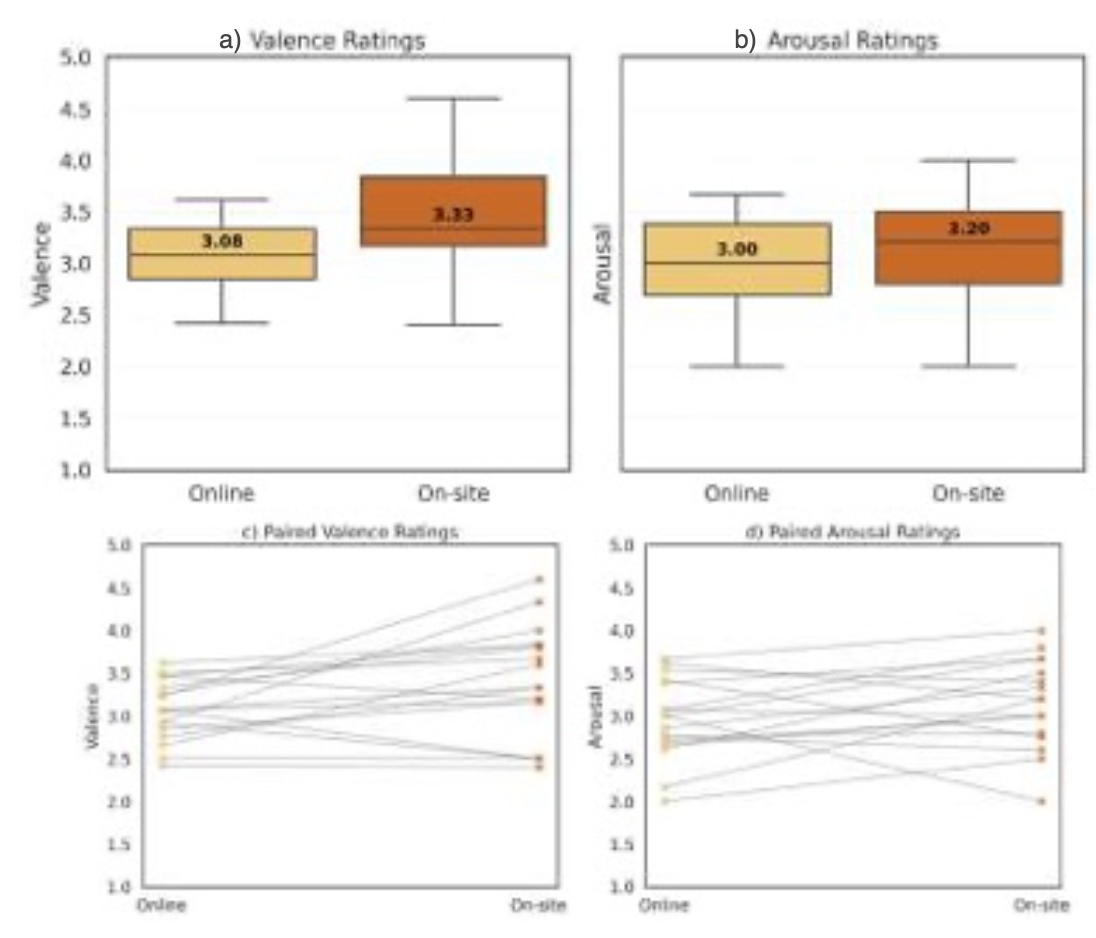}
\caption{Comparison between online and on-site 
affective ratings. Paired plots for valence and 
arousal showing individual changes across 15 
fa\c{c}ades, with corresponding boxplots 
displaying central tendency and dispersion of 
scores. Valence ratings were significantly 
higher in the on-site condition relative to 
online evaluations, while arousal ratings showed 
no significant difference between conditions. 
Source: Author.\label{fig26b}}
\end{figure}
\unskip
\vspace{6pt}

\textit{Cross-context affective consistency.} 
Affective responses showed the lowest 
cross-context reliability of the three 
measurement layers. Valence ratings 
differed significantly between online 
and in-situ conditions (paired $t$-test, 
$p = 0.024$; ICC $= 0.332$), while 
arousal ratings did not differ 
significantly ($p = 0.183$) but also 
demonstrated limited agreement 
(ICC $= 0.364$). These patterns indicate 
that affective responses, particularly 
valence, are substantially shaped by 
multisensory, contextual, and embodied 
factors not fully captured in image-based 
evaluations, consistent with broader 
evidence that affect is highly 
context-dependent \citep{Barrett2017, 
Gregorians2022}.

Taken together, the validation results 
reveal a hierarchically decreasing 
stability across measurement layers: 
physical features show partial alignment 
with human perception; perceptual 
attributes exhibit moderate cross-context 
consistency; and affective responses 
display the greatest sensitivity to 
viewing context. This structure 
reinforces the necessity of perceptual 
intermediaries in predictive models of 
architectural affect and highlights 
the bounded ecological validity of 
image-based affective evaluation.

\section{Discussion}

\subsection{7.1 Results and Relevance}

Human perception constitutes the critical interface between façade features and affective response. This study provides empirical evidence that the attributes of the human perception of the façade attributes of complexity, transparency, and materiality, systematically shape affective responses, operationalized as arousal and valence. Across a multi-layered analytical framework, the findings establish a theoretically grounded linkage between façade composition and affective experience, advancing the empirical basis of how architectural interfaces modulate affect.

The study contributes a dataset, a pipeline, and an empirical structure of relationships. First, it introduces the Cambridge Façade Affect Dataset (CFAD), a high-resolution, affectively annotated façade dataset designed for both computational and perceptual research. Second, it develops a replicable computational–perceptual pipeline that integrates machine-extracted features with human response data, enabling scalable analysis of façade-level affective impact. Third, it systematically reveals the correlational structure, non-linear dynamics, and mediating relationships between façade attributes and affective responses, providing an empirical foundation for future causal modelling and theory-building in affective urbanism.

Perceived façade features—not computational descriptors—drive affective outcomes. The results demonstrate that human-perceived façade characteristics, particularly complexity and materiality, are robust predictors of affect. Linear mixed-effects models show that perceived complexity increases arousal, while perceived material naturalness enhances valence, aligning with established psychophysical and neuroarchitectural accounts of environmental preference \citep{Coburn2020}. However, beyond confirming these associations, the findings reveal a deeper epistemic condition: complexity does not operate as a scalar variable but as a relational construct. As indicated by SHAP-based analysis, identical levels of measured complexity may yield divergent affective responses depending on co-present perceptual conditions. Architectural experience, therefore, cannot be reduced to monotonic mappings between physical metrics and affect.

Affective meaning is constructed through perceptual interpretation. This tension becomes more pronounced when contrasting computational and perceptual predictors. Machine-derived features such as edge density and WWR exhibit limited explanatory power, whereas human ratings capture latent interpretive dimensions that likely involve semantic recognition, spatial coherence, and embodied familiarity. In this sense, perception is not a passive registration of stimuli but an active cognitive–affective synthesis. Complexity does not merely stimulate; it structures attention and invites interpretation. Materiality does not simply register texture; it activates associative memory and embodied expectations.

Affect is context-sensitive, while perception is comparatively stable. A further contribution of this study lies in identifying an asymmetry between perceptual and affective stability across contexts. While perceived complexity and materiality demonstrate moderate consistency between online and in-situ conditions, valence exhibits significant variation, indicating strong context dependence. This suggests that perceptual judgments may be relatively robust to representational changes, whereas affective responses are more sensitive to atmospheric, multisensory, and situational factors.

Toward a perceptually mediated model of architectural affect. Taken together, these findings support a human-centred model of architectural cognition in which perceptual judgments function as epistemic mediators between physical form and affective meaning. Rather than treating perceptual variability as noise, this study positions it as the mechanism through which architectural features acquire affective significance. In this model, façades do not directly encode emotion; they provide structured stimuli that are interpreted, negotiated, and ultimately experienced through perception.

\subsection{7.1 Results and Relevance}

Affective responses to façades are polarised between perceptual engagement and affective invisibility. The image-based evaluation reveals a clear divergence in emotional responses to architectural façades, with certain typologies consistently perceived as either affectively flat or emotionally engaging. Buildings receiving the lowest scores in both arousal and valence, such as Image 80 (Department of Physics, Kapitza Building) were characterised by visual simplicity, material ambiguity, and low perceptual salience. Averaged across participant ratings, this façade yielded an arousal score of 1.75/5 and a valence score of 2.00/5, indicating not active dislike, but rather a lack of affective registration.

Participants frequently described such façades as ``plain'', ``cold'', or ``uninteresting''. Architecturally, these typologies resemble generic mid-rise office blocks, with minimal articulation, planar curtain walls, and repetitive fenestration. Crucially, they lack perceptual differentiation, rhythmic variation, and material contrast. This absence of structured visual stimuli aligns with arousal theory \citep{Berlyne1971}, whereby low novelty and limited informational richness suppress perceptual activation. In parallel, material ambiguity, where natural and artificial surfaces are visually indistinguishable, contributes to what can be described as affective neutrality or spatial anonymity.

Affective engagement emerges from structured complexity and material legibility. In contrast, façades such as Image 55 (St John’s College New Court) occupy the upper range of both arousal (4.50/5) and valence (4.33/5). These buildings exhibit layered compositional complexity, strong rhythmic articulation, and materially legible surfaces, often incorporating natural materials such as stone, brick, and timber. Importantly, affective engagement does not derive from any single attribute in isolation, but from the coherent organisation of multiple perceptual cues, including depth, variation, and material expression.

Regression analyses further support this interpretation. A significant three-way interaction between perceived complexity, transparency, and materiality predicts both arousal ($\beta = 0.126$, $p = 0.020$) and valence ($\beta = 0.129$, $p = 0.014$), indicating that affective responses are driven by combined feature configurations rather than independent variables. This suggests that façade experience is integrative rather than additive, emerging from the interplay between structural richness, visual permeability, and material character.

“Boring” façades are not negatively evaluated, but affectively unregistered. A key insight of this study is that low-performing façades are not necessarily perceived as unpleasant, but rather as affectively negligible. These buildings occupy a low-arousal, low-valence region of the affective space, indicating minimal emotional engagement rather than negative appraisal. In this sense, the absence of affect constitutes a distinct category of architectural experience - one defined not by aversion, but by perceptual indifference.

Affective resonance depends on perceptual legibility and compositional coherence. By contrast, façades that exhibit clear compositional logic, material authenticity, and perceptual richness are more likely to elicit strong affective responses. This finding aligns with theories of aesthetic fluency \citep{Reber2004}, which propose that perceptual pleasure arises from stimuli that are both structured and interpretable. In this framework, architectural façades generate affect not through complexity alone, but through the legible organisation of complexity into coherent visual patterns.

Taken together, these results suggest that affective impact in architecture is not determined by the presence of visual features per se, but by their perceptual organisation and interpretability. Façades that fail to establish this relationship remain affectively invisible, while those that do achieve perceptual coherence become emotionally resonant.

\section{Limitations}

The correspondence between the machine-derived and 
human-perceived attributes fa\c{c}ade was partial 
and attribute-dependent, with materiality showing 
meaningful cross-modal alignment ($\rho = 0.431$) 
while complexity exhibited only weak correspondence 
($\rho = 0.127$, $p = 0.243$). This divergence 
reflects a domain mismatch inherent to contemporary 
computer vision pipelines: models pre-trained on 
general-purpose, object-centric datasets encounter 
the historically layered, ornamentally rich, and 
compositionally irregular fa\c{c}ades characteristic 
of the Cambridge context with insufficiently 
specified representational vocabularies 
\citep{Luddecke2022, LiuV2022}. Perceived 
complexity, in particular, appears to recruit 
higher-order structural and semantic information, 
including hierarchical organisation, perceptual 
grouping, and typological legibility, that is 
not adequately captured by edge-based descriptors 
alone \citep{Doersch2012, Marr1980}.

A related limitation concerns the spatial 
assumptions embedded in current pipelines. Edge 
density, window-to-wall ratio, and material 
classification are computed as globally averaged 
statistics across the fa\c{c}ade image, implicitly 
treating the surface as a spatially uniform field. 
Human perceptual attention, however, is inherently 
non-uniform: affective responses to architecture 
are frequently anchored in localised features, 
such as a material transition at an entrance, a 
rhythmically articulated window sequence, or a 
zone of ornamental density, that carry 
disproportionate emotional weight relative to 
their spatial extent \citep{Beder2024a, 
Hollander2019EyeTracking}. This 
perceptual--epistemic misalignment suggests that 
predictive models of architectural affect may 
benefit substantially from saliency-weighted or 
attention-modulated representations, in which 
regions of high perceptual salience are assigned 
greater influence in feature aggregation 
\citep{LiuV2022}.

The cross-context affective validation further 
reveals a ceiling on the ecological validity of 
image-based evaluation. Valence ratings differed 
significantly between online and in-situ conditions 
($p = 0.024$, ICC $= 0.332$), indicating that 
perceived pleasantness is substantially shaped by 
multisensory, embodied, and contextual factors 
that static orthographic images cannot fully 
convey \citep{Barrett2017, Gregorians2022}. 
Static image paradigms therefore provide a 
tractable but bounded approximation of in-situ 
affective experience, and findings derived from 
them should be interpreted with this constraint 
in view.

Finally, the present study is geographically 
bounded to Cambridge, a city characterised by 
a specific and historically coherent architectural 
vocabulary. The dataset's stylistic range, while 
diverse within this setting, does not represent 
the full spectrum of contemporary global 
fa\c{c}ade typologies, and the perceptual and 
affective relationships documented here may not 
generalise without qualification to other urban 
contexts.

These limitations define a clear forward 
trajectory. Developing architecturally specific, 
perceptually labelled training datasets, along 
the lines of the CFAD introduced here, would 
enable perception-informed retraining of vision 
models in which human judgements serve as 
supervisory signals rather than post-hoc 
validation targets \citep{Kosti2020}. Integrating 
saliency-weighted processing and 
attention-modulated representations into 
fa\c{c}ade analysis pipelines \citep{LiuV2022} 
would address the local--global misalignment 
identified here. Extending evaluation to 
immersive, physiologically instrumented paradigms, 
including virtual reality coupled with mobile 
EEG and electrodermal sensing, would enable 
affective responses to be captured as temporal 
trajectories during spatial navigation rather 
than as static image ratings \citep{Chiamulera2022, 
Banaei2017}. Replication across architecturally, 
climatically, and culturally distinct urban 
environments will be necessary to establish the 
cross-contextual generalisability of the 
perceptual--affective model developed here, and 
to build the large-scale, ecologically grounded 
datasets that perception-informed urban AI will 
ultimately require.

\authorcontributions{Conceptualization, C.W. 
and M.G.-M.; methodology, C.W. and H.D.; 
software, C.W.; validation, C.W., H.D. and 
M.G.-M.; formal analysis, C.W.; investigation, 
C.W.; resources, C.W.; data curation, C.W.; 
writing - original draft preparation, C.W.; 
writing - review and editing, C.W., H.D. and 
M.G.-M.; visualization, C.W.; supervision, 
M.G.-M.; project administration, M.G.-M. All 
authors have read and agreed to the published 
version of the manuscript.}

\funding{This research received no external 
funding.}

\institutionalreview{The study was conducted 
in accordance with the Declaration of Helsinki, 
and approved by the Department of Architecture 
Research Ethics Committee at the University of 
Cambridge (protocol code AHA/PG/ST24100021).}

\informedconsent{Informed consent was obtained 
from all subjects involved in the study.}

\dataavailability{The Cambridge Fa\c{c}ade 
Affect Dataset (CFAD), including fa\c{c}ade 
images, machine-derived feature scores, and 
anonymised participant ratings, will be made 
publicly available upon acceptance of this 
manuscript. Data requests prior to publication 
may be directed to the corresponding author.}

\acknowledgments{The authors wish to thank all 
participants who contributed their time to the 
perceptual surveys. The field-based validation 
study was conducted within the University of 
Cambridge. No generative AI tools were used in 
drafting any aspect of this manuscript.}

\conflictsofinterest{The authors declare no 
conflicts of interest.}

\abbreviations{Abbreviations}{
The following abbreviations are used in this 
manuscript:\\

\noindent
\begin{tabular}{@{}ll}
CFAD  & Cambridge Fa\c{c}ade Affect Dataset \\
SAM   & Self-Assessment Manikin \\
WWR   & Window-to-Wall Ratio \\
LME   & Linear Mixed-Effects Model \\
ICC   & Intraclass Correlation Coefficient \\
BIBD  & Balanced Incomplete Block Design \\
CV    & Computer Vision \\
EEG   & Electroencephalography \\
GSR   & Galvanic Skin Response \\
VR    & Virtual Reality \\
\end{tabular}
}


\reftitle{References}
\bibliography{ref}

\end{document}